# Astrophysical S-factors of radiative capture of protons on $^2$H, $^3$H, $^6$Li and $^{12}$C nuclei

S.B. Dubovichenko, A.V. Dzhazairov-Kakhramanov

*V.G. Fessenkov's Astrophysical Institute, 050022, Almaty, Kazakhstan*
*E-mail address:* sergey@dubovichenko.ru, albert-j@yandex.ru

**Abstract**

We have considered the possibility to describe the astrophysical *S*-factors of some light atomic nuclei on the basis of the potential cluster model by taking into account the supermultiplet symmetry of wave functions and splitting the orbital states according to Young's schemes. Such an approach allows analyzing the structure of inter-cluster interactions, to determine the presence of allowed and forbidden states in the interaction potential and so the number of nodes of the orbital wave function of cluster relative motion.

## 1. Introduction

The experimental data on cross-sections of nuclear reactions and their analysis within various theoretical models are the major source of information about nuclear structure, nature and mechanism of nucleus-nucleus (cluster-cluster) interaction. Researches in nuclear astrophysics are complicated since in many cases only theoretical predictions can supply deficient experimental information on characteristics of nuclear reactions. This difficulty is associated with the low energy of interaction of matter in the stars which ranges from tenth to tens keV.

Thus, it is impossible to measure directly the cross-sections required for astrophysical calculations of nuclear reactions. Usually, the cross-sections are measured at higher energies (about hundreds of keV) and then extrapolated into the energy range representing interest for nuclear astrophysics [1]. However, a simple extrapolation of experimental data into the astrophysical range is not always correct due to the fact that the experimental measurements of thermonuclear cross-sections are carried out at rather high energies (0.2-1.0 MeV). The experimental error band of the determination of the astrophysical *S*-factor at the energy range of 10-300 keV in different cluster systems can reach up to ±100% extremely lowering the value of such an extrapolation. In this situation the role of theoretical calculations becomes considerably more important.

The calculations carried out on the basis of the chosen theoretical concepts are compared with the existing experimental data (at the energy ranges where these data are available). This allows making certain conclusions about the quality of the physical model used and thus to select concepts and approaches leading to the best agreement with the experiment, which means that they best describe the real situation in the atomic nucleus at these energies. Then on the basis of the chosen model one can make the calculations in the astrophysical energy range, and this is not a simple extrapolation of the experimental data, because of the fact that such an approach has quite a definite microscopic rationale.

The considered potential cluster model (PCM) of atomic nucleus on the basis of calculated potentials of nuclear interaction allows calculating the required nuclear characteristics, such us the cross-sections of different photoreactions and the astrophysical *S*- factors, quite easily. The approach used here for the theoretical analysis of such characteristics allows us to obtain results at the lowest energies (down to 1 keV).

We choose the potential cluster model because the probability of formation of nucleonic associations is comparatively high in light atomic nuclei. This is evidenced by the numerous experimental data and theoretical results, which have been received over the past fifty years [2]. The role of exchange effects between clusters is comparatively small at the range of low energies and momentums. This fact allows us using the simple potential cluster model without antisymmetrization of wave functions for thermonuclear reactions and for any nuclear processes at low energies on light nuclei, too.

Thus, the one-channel potential cluster model is a good approximation to the situation really existing in the atomic nucleus in many cases and for various light nuclei. Such a model allows making any calculations of nuclear characteristics in the scattering processes and bound states quite easily, even in the systems where many-body problem solution methods are very cumbersome in the digital implementation or do not lead to certain numerical results at all.

For example, on the basis of measurements of the differential cross-sections of the elastic scattering of nuclear particles [3] and [4] it is possible to perform the phase shift analysis, which at energies lower than 1 MeV usually includes the $S$-wave only. The data of differential cross-sections usually measured at 10-15 degrees of scattering at the required energy range allow us carrying out the most complete and accurate phase shift analysis and receiving the phase shifts of elastic scattering. Further, we can construct the nuclear potential of the inter-cluster interaction using the received phase shifts of scattering. This potential, in its turn, allows us to make any calculations of nuclear processes, for example, the calculation of the astrophysical $S$-factor of radiative capture at low energies, which is of interest for nuclear astrophysics.

We considered the astrophysical $S$-factors on the basis of the PCM which takes into account the supermultiplet symmetry of wave functions (WF) and the splitting of orbital states according to Young's schemes. This approach allows us analysing the structure of inter-cluster interactions, identifying allowed states (AS) and forbidden states (FS) in the interaction potential, and thus, the number of WF nodes of cluster relative motion [5,6].

## 2. Methods

### 2.1. Astrophysical S-factors

The formula for the astrophysical S-factor of the radiative capture process is of the form [7]

$$S(EJ) = \sigma(EJ) E_{cm} \exp\left(\frac{31.335 \, Z_1 Z_2 \sqrt{\mu}}{\sqrt{E_{cm}}}\right), \quad (1)$$

where $\sigma$ is the total cross-section of the radiative capture (barn), $E_{cm}$ is the center-of-mass energy of particles (keV), $\mu$ is the reduced mass of input channel particles at the radiative capture (atomic mass unit) and $Z_{1,2}$ are the particle charges in elementary charge units. The numerical coefficient 31.335 was received on the basis of up-to-date values of fundamental constants, which are given in [8].

The total cross-sections of radiative capture for electric $EJ(L)$ transitions, caused by the orbital part of electric operator, in a cluster model are given, for example, in works [9] or [10] and may be written as

$$\sigma(E) = \sum_{J,J_f} \sigma(EJ, J_f), \quad (2)$$

$$\sigma(EJ, J_f) = \frac{8\pi K e^2}{\hbar^2 q^3} \frac{\mu}{(2S_1+1)(2S_2+1)} \frac{J+1}{J[(2J+1)!!]^2} A_J^2(K) \sum_{L_i, J_i} |P_J(EJ, J_f) I_J|^2,$$

where

$$P_J(EJ, J_f) = \delta_{S_i S_f} (-1)^{J_i + S + L_f + J} \sqrt{(2J+1)(2L_i+1)(2J_i+1)(2J_f+1)} (L_i 0 J 0 | L_f 0) \begin{Bmatrix} L_i & S & J_i \\ J_f & J & L_f \end{Bmatrix},$$

$$A_J(K) = K^J \mu^J \left(\frac{Z_1}{M_1^J} + (-1)^J \frac{Z_2}{M_2^J}\right),$$

$$I_J = \langle L_f J_f | R^J | L_i J_i \rangle. \quad (3)$$

Here, $\mu$ is the reduced mass and $q$ is the wave number of input channel particles; $L_f$, $L_i$, $J_f$, $J_i$ are particle momenta for input ($i$) and output ($f$) channels; $S_1$, $S_2$ - spins; $M_{1,2}$, $Z_{1,2}$, are masses and charges of input channel particles (1 or 2); $K^J$, $J$ - the wave number and the momentum of γ-quanta; $I_J$ is the integral taken over wave functions of initial and final states, that is functions of the relative cluster motion with the intercluster distance $R$. Sometimes, the spectroscopic factor $S_{Jf}$ of the final state is used in the given formulas for cross-sections, but it is equal to one in the potential cluster model that we used, as it was in work [9].

Using the formula from [11] for the magnetic transition $M1(S)$ caused by the spin part of the magnetic operator we can obtain

$$P_1(M1) = \delta_{S_i S_f} \delta_{L_i L_f} (-1)^{S + J_f + L + 1} \sqrt{S(S+1)(2S+1)(2J_i+1)(2J_f+1)} \begin{Bmatrix} S & L & J_i \\ J_f & 1 & S \end{Bmatrix},$$

$$A_1(K) = i\frac{\hbar K}{m_0 c}\sqrt{3}[\mu_1 + \mu_2], \tag{4}$$

$$I_1 = \langle \Phi_f | \Phi_i \rangle,$$

where $\mu_1$ and $\mu_2$ are magnetic momenta of proton and $^2$H, which are taken from work [12] ($\mu_H$=0.857 and $\mu_p$=2.793).

*2.2. Potentials and functions*

Potentials of intercluster interactions with a point-like Coulomb potential are represented as

$$V(R)=V_0\exp(-\alpha R^2)+V_1\exp(-\gamma R) \tag{5}$$

or

$$V(R)=V_0\exp(-\alpha R^2). \tag{6}$$

The expansion of WF of relative cluster motion in nonorthogonal Gaussian basis and the independent variation of parameters [10] are used in the variational method (VM)

$$\Phi_L(R) = \frac{\chi_L(R)}{R} = R^L \sum_i C_i \exp(-\beta_i R^2), \tag{7}$$

where $\beta_i$ and $C_i$ are the variational expansion parameters and expansion coefficients.

The behavior of the wave function of bound states (BS) at long distances is characterized by the asymptotic constant $C_W$, having a general form [13]

$$\chi_L(R) = \sqrt{2k_0}\, C_W W_{\eta L}(2k_0 R), \tag{8}$$

where $\chi_L$ is the numerical wave function of the bound state obtained from the solution of the radial Schrödinger equation and normalized to unity; $W$ is the Whittaker function of the bound state which determines the asymptotic behavior of the WF and represents the solution of the same equation without nuclear potential, i.e. long distance solution; $k_0$ is the wave number determined by the channel bound energy; $\eta$ is the Coulomb parameter; $L$ is the orbital momentum of the bound state.

The root-mean-square mass radius is represented as

$$R_m^2 = \frac{M_1}{M}\langle r_m^2 \rangle_1 + \frac{M_2}{M}\langle r_m^2 \rangle_2 + \frac{M_1 M_2}{M^2} I_2,$$

where $M_{1,2}$ and $\langle r_m^2 \rangle_{1,2}$ are the masses and squares of mass radii of clusters, $M=M_1+M_2$, $I_2$ - the integral

$$I_2 = \langle \chi_L(R) | R^2 | \chi_L(R) \rangle$$

of the *R* inter-cluster distance and the integration is over radial WF $\chi_L(R)$ of cluster relative motion with the orbital momentum *L* (3).

The root-mean-square charge radius is represented as

$$R_z^2 = \frac{Z_1}{Z}\langle r_z^2 \rangle_1 + \frac{Z_2}{Z}\langle r_z^2 \rangle_2 + \frac{(Z_2 M_1^2 + Z_1 M_2^2)}{ZM^2} I_2,$$

where $Z_{1,2}$ and $\langle r_z^2 \rangle_{1,2}$ are the charges and squares of charge radii of clusters, $Z=Z_Z+Z_2$, $I_2$ - the abovementioned integral.

The wave function $\chi_L(R)$ or $|L_i J_i\rangle$ is the solution of the radial Schrödinger equation of the form

$$\chi''_L(R) + [k^2 - V(R) - V_c(R) - L(L+1)/R^2]\chi_L(R) = 0,$$

where *V(R)* is the inter-cluster potential represented as (5) or (6) (dim. fm$^{-2}$); $V_c(R)$ is the Coulomb potential; *k* is the wave number determined by the energy *E* of interaction particles $k^2 = 2\mu E/\hbar^2$; μ is the reduced mass.

The states with the minimal spin in the scattering processes of some light atomic nuclei are "mixed" according to orbital Young's schemes, for example the doublet p$^2$H state [5] is mixed according to schemes {3} and {21}. At the same time, the bound forms of these states are "pure" according to Young's schemes, for example, the doublet p$^2$H channel of the $^3$He nucleus is "pure" according to scheme {3}. The method of splitting of such states according to Young's schemes is suggested in works [2, 5] where in all cases the "mixed" phase shift of scattering can be represented as a half-sum of "pure" phase shifts {f$_1$} and {f$_2$}

$$\delta^{\{f_1\}+\{f_2\}} = 1/2\left(\delta^{\{f_1\}} + \delta^{\{f_2\}}\right) \tag{9}$$

In this case it is considered that {f$_1$}={21} and {f$_2$}={3}, and the doublet phase shifts, derived from the experiments, are "mixed" in accordance with these two Young's schemes. If we suppose that instead of the "pure" quartet phase shift with the symmetry {21} one can use the "pure" doublet phase shift of p$^2$H scattering with same symmetry, then it is easy to find the "pure" doublet p$^2$H phase shift with {3} symmetry [5] and use it for the construction of the interaction potential "pure" according to Young's schemes. The latter can be used for the description of the characteristics of the bound state. In this case such a potential allows us to consider the bound p$^2$H state of the $^3$He nucleus. Similar ratios apply to other light nuclear systems as well, and in each specific case we will analyze the AS and FS structure for both the scattering potentials and the interactions of the ground bound states.

*2.3. Phase shift analysis*

Using experimental data of differential cross-sections of scattering, it is possible to find a set of phase shifts $\delta_{S,L}^J$, which can reproduce the behavior of these cross-sections with certain accuracy. Quality of description of experimental data on the basis of a certain theoretical function (functional of several variables) can be estimated by the $\chi^2$ method which is written as

$$\chi^2 = \frac{1}{N}\sum_{i=1}^{N}\left[\frac{\sigma_i^t(\theta)-\sigma_i^e(\theta)}{\Delta\sigma_i^e(\theta)}\right]^2 = \frac{1}{N}\sum_{i=1}^{N}\chi_i^2, \qquad (10)$$

where $\sigma^e$ and $\sigma^t$ are experimental and theoretical (i.e. calculated for some defined values of phase shifts $\delta_{S,L}^J$ of scattering) cross-sections of elastic scattering of nuclear particles for $i$-angle of scattering, $\Delta\sigma^e$ – the error of experimental cross-sections at these angles, $N$ – the number of measurements.

The less $\chi^2$ value, the better description of experimental data on the basis of the chosen phase shift of scattering set. Expressions describing the differential cross-sections represent the expansion of some functional $d\sigma(\theta)/d\Omega$ to the numerical series and it is necessary to find such variational parameters of expansion $^{2,4}\delta_L$ which are the best for the description of its behavior. Since the expressions for the differential cross-sections are exact, then as $L$ approaches infinity the value of $\chi^2$ must vanish to zero. This criterium is used for choosing a certain set of phase shifts ensuring the minimum of $\chi^2$ which could possibly be the global minimum of a multiparameter variational problem [14].

So, for example, for p$^6$Li system in order to find nuclear phase shifts of scattering using experimental cross-sections, the procedure of minimization of the functional $\chi^2$ as a function of $2L+2$ variables, each of which is a phase shift $^{2,4}\delta_L$ of a certain partial wave without spin-orbital splitting, was carried out. To solve this problem we serched for the minimum of $\chi^2$ within a limited range of values for such variables. But it is possible to find a lot of local minima of $\chi^2$ with the value of about one in this range. Choosing the smallest of them allows hoping that this minimum will correspond to the global minimum which is a solution of this variational problem. Then, the value of this minimum should decrease more or less smoothly as the number of partial waves increases. We used these criteria and methods for the phase shift analyses in the p$^6$Li and p$^{12}$C systems at low energies – the systems important for the astrophysical calculations.

The exact mass values of the particles were taken for all our calculations [12], and the $\hbar^2/m_0$ constant was taken to be 41.4686 MeV fm$^2$. The Coulomb parameter $\eta=\mu Z_1Z_2 e^2/(q\hbar^2)$ was represented as $\eta = 3.44476\ 10^{-2} Z_1Z_2 \mu/q$, where $q$ is the wave number determined by the energy of interacting particles in the input channel (in fm$^{-1}$), $\mu$ - the reduced mass of the particles (atomic mass unit), $Z$ - the particle charges in elementary charge units. The Coulomb potential with $R_c=0$ was represented as $V_c$ (MeV)$=1.439975\ Z_1Z_2/r$, where $r$ is the distance between the input channel particles (fm).

# 3. p²H radiative capture

The first process under consideration is the radiative capture

p+²H→³He+γ ,

which is a part of hydrogen cycle and gives a considerable contribution to energy efficiency of thermonuclear reactions [15] accounting for burning of the Sun and stars of our Universe. The potential barrier for interacting nuclear particles of the hydrogen cycle is the lowest. Thus, it is the first chain of nuclear reactions which can take place at ultralow energies and star temperatures.

For this chain, the process of the radiative p²H capture is the basic process for the transition from the primary proton fusion

p+p→²H+e⁻+ν$_e$

to the capture reaction of two ³He nuclei [16], which is one of the final processes

³He+³He→⁴He+2p

in the p-p-chain.

The theoretical and experimental study of the radiative p²H capture in detail is of fundamental interest not only for nuclear astrophysics, but also for nuclear physics of ultralow energies and lightest atomic nuclei [17]. That is why the experimental researches into this reaction are in progress and a short time ago the new experimental data in the range down to 2.5 keV appeared.

*3.1. Potentials and phase shifts of scattering*

Earlier, the total cross sections of the photoprocesses of lightest ³He and ³H nuclei were considered in the frame of the potential cluster model in our work [6]. E1 transitions resulting from the orbital part of the electric operator $Q_{Jm}(L)$ [10] were taken into account in these calculations of the photodecays of ³He and ³H nuclei into p²H and n²H channels. The values of E2 cross-sections and cross-sections depending on the spin part of the electric operator turned out to be several orders less.

Further, it was assumed that E1 electric transitions in N²H system are possible between ground "pure" (scheme {3}) ²S state of ³H and ³He nuclei and doublet ²P scattering state mixed according to Young's schemes {3}+{21} [17]. On the basis of the approach used it was possible to obtain quite reasonable results describing the experimental data of ³H and ³He nuclei photodecay into the cluster channels [6].

To calculate photonuclear processes in the systems under consideration the nuclear part of the potential of inter-cluster p²H and n²H interactions is represented as (5) with a point-like Coulomb potential, $V_0$ - the Gaussian attractive part, and $V_1$ - the exponential repulsive part. The potential of each partial wave was constructed so as to correctly describe the respective partial phase shift of the elastic scattering [18]. Using this concept, the potentials of the p²H interaction of the scattering processes were received. The parameters of such potentials were fully given in works [6,10,19], and parameters for doublet scattering states mixed according to Young's schemes are listed in Table 1.

Table 1
Potentials of the p$^2$H [6] interaction in the doublet channel

| $^{2S+1}L$, {$f$} | $V_0$ (MeV) | $\alpha$ (fm$^{-2}$) | $V_1$ (MeV) | $\gamma$ (fm$^{-1}$) |
|---|---|---|---|---|
| $^2S$, {3} | -34.76170133 | 0.15 | --- | --- |
| $^2P$, {3}+{21} | -10.0 | 0.16 | +0.6 | 0.1 |
| $^2S$, {3}+{21} | -35.0 | 0.1 | --- | --- |

Then, in the doublet channel mixed according to Young's schemes {3} and {21} [5], the "pure" phases with scheme {3} were separated (9) and on their basis the "pure" $^2S$ potential of the bound state of the $^3$He nucleus in the p$^2$H channel was constructed [6,10,19].

The calculations of the E1 transition [6] show that the best results for the description of the total cross-sections of the $^3$He nucleus photodecay for the γ-quanta energy range 6-28 MeV, including the maximum value at Eγ=10-13 MeV, can be found if we use the potentials with peripheric repulsion of the $^2P$-wave of the p$^2$H scattering (table 1) and the "pure" according to Young's schemes $^2S$-interaction of the bound state (BS) of the Gaussian form (5) with parameters

$V_0$ = -34.75 MeV, $\alpha$ = 0.15 fm$^{-2}$, $V_1$ = 0 ,

which were obtained, primarily, on the basis of the correct description of the bound energy (with the accuracy up to few keV) and the charge radius of the $^3$He nucleus. The calculations of the total cross-sections of the radiative p$^2$H capture and astrophysical S-factors were made with these potentials at the energy range down to 10 keV [6-10]. Though, at that period of time we only knew S-factor experimental data in the range above 150-200 keV [20].

Recently, the new experimental data on the p$^2$H S-factor in the range down to 2.5 keV appeared in [21-23]. That is why, it is interesting to know if it is possible to describe the new data on the basis of the E1 and M1 transitions in the potential cluster model with the earlier obtained $^2P$-interaction of scattering and $^2S$-potential of the bound p$^2$H state adjusted in this work.

Our preliminary results have shown that for the S-factor calculation at the energy range of about 1 keV it is necessary to improve the accuracy of finding the bound energy of the p$^2$H system in the $^3$He nucleus. It must be better than 1-2 keV [6]. The behaviour of the tail of the wave function of the bound state should be controlled more strictly at long distances. Then, it is necessary to improve the accuracy of finding Coulomb wave functions which determine the asymptotic behaviour of the scattering WF in the $^2P$-wave.

The parameters of the "pure" doublet $^2S$-potential according to Young's scheme {3} were adjusted using opportunities of a new computer programs based on the finite-difference method (FDM) for a more accurate description of the experimental bound energy of $^3$He nuclei in p$^2$H channel. This potential (Table 1) has become somewhat deeper than the potential we used in our work [6] and leads to a total agreement between calculated -5.4934230 MeV and experimental -5.4934230 MeV bound energies, which is obtained by using the exact mass values of particles [12]. The difference between potentials given in work [6] and in Table 1 is primarily due to using the exact mass values of particles and more accurate description of the $^3$He nucleus bound energy in the p$^2$H channel. For these computations the absolute accuracy of searching for the bound energy in our computer program based on the finite-difference method was taken to be at the level of 10$^{-8}$ MeV.

The value of the $^3$He charge radius with this potential equals 2.28 fm, which is a little higher than the experimental values listed in Table 2 [12,24,25]. The experimental radii of proton and deuteron, which are also given in Table 2, are used for these calculations and the latter is larger than the radius of the $^3$He nucleus. Thus, if the deuteron is present in the $^3$He nucleus as a cluster, it must be compressed by about 20-30% of its size in free state for a correct description of the $^3$He radius [10].

Table 2
Experimental masses and charge radii of light nuclei used in these calculations [12, 24, 25]

| Nucleus | Radius, (fm) | Mass |
|---|---|---|
| p | 0.8768(69) | 1.00727646677 |
| $^2$H | 2.1402(28) | 2.013553212724 |
| $^3$H | 1.63(3); 1.76(4); 1.81(5) <br> The average value is 1.73 | 3.0155007134 |
| $^3$He | 1.976(15); 1.93(3); 1.877(19); 1.935(30) <br> The average value is 1.93 | 3.0149322473 |
| $^4$He | 1.671(14) | 4.001506179127 |

The asymptotic constant $C_W$ with Whittaker asymptotics (8) [26] was calculated for controlling behavior of WF of BS at long distances; its value in the range of 5-20 fm equals $C_W$=2.333(3). The error given here is found by averaging the constant in the range mentioned above. The experimental data known for this constant give the values of 1.76-1.97 [27,28], which is slightly less than the value obtained here. It is possible to give results of three-body calculations [29], where a good agreement with the experiment [30] for the ratio of asymptotic constants of $^2S$ and $^2D$ waves was obtained and the value of the constant of $^2S$ wave was found to be $C_W$=1.878.

But in work [13], which is more recent than [27, 28], the value of 2.26(9) is given for the asymptotic constant, and this is in a good agreement with our calculations. One can see from the considerable data that there is a big difference between the experimental results of asymptotic constants received in different periods. These data are in the range from 1.76 to 2.35 with the average value of 2.06.

In the cluster model the value of $C_W$ constant depends significantly on the width of the potential well and it is always possible to find other parameters of $^2S$-potential of BS, for example:

$V_0 = - 48.04680730$ MeV и $\alpha = 0.25$ fm$^{-2}$, (11)
$V_0 = - 41.55562462$ MeV и $\alpha = 0.2$ fm$^{-2}$, (12)
$V_0 = - 31.20426327$ MeV и $\alpha = 0.125$ fm$^{-2}$, (13)

which give the same value of the bound energy of $^3$He in p$^2$H channel. The first of them at distances of 5-20 fm leads to asymptotic constant $C_W$=1.945(3) and charge radius $R_{ch}$=2.18 fm, the second variant gives $C_W$=2.095(5) and $R_{ch}$=2.22 fm, the third variant - $C_W$=2.519(3) and $R_{ch}$=2.33 fm.

It can be seen from these results that the potential (11) allows obtaining the charge radius the closest to the experimental values. Further reduction of the potential width could give a more accurate description of its value, but, as it will be shown later, will not allow us to describe the $S$-factor of the p$^2$H capture. In this sense, the slightly wider potential (12) has the minimal acceptable width of the potential well which leads to asymptotic constant almost equal to its experimental average value 2.06 and gives a possibility to describe quite well the astrophysical *S*-factor in a wide energy range.

The variational method is used for an additional control of the accuracy of bound energy calculations for the potential from Table 1, which allowed obtaining the bound energy of -5.4934228 MeV by using independent variation of parameters and the grid having dimension 10. The asymptotic constant $C_W$ of the variational WF at distances of 5-20 fm remains at the level of 2.34(1). The variational parameters and expansion coefficients of the radial wave function for this potential having form (7) are listed in Table 3.

Table 3
The variational parameters and expansion coefficients of the radial WF of the bound state of the p$^2$H system for the potential from Table 1. The normalization of the function with these coefficients in the range 0-25 fm equals N=0.999999997

| *i* | $\beta_i$ | $C_i$ |
|---|---|---|
| 1 | 2.682914012452794E-001 | -1.139939646617903E-001 |
| 2 | 1.506898472480031E-002 | -3.928173077162038E-003 |
| 3 | 8.150892061325998E-003 | -2.596386495718163E-004 |
| 4 | 4.699184204753572E-002 | -5.359449556198755E-002 |
| 5 | 2.664477374725231E-002 | -1.863994304088623E-002 |
| 6 | 4.4687619986542310E+001 | 1.098799639286601E-003 |
| 7 | 8.482112461789261E-002 | -1.172712856304303E-001 |
| 8 | 1.541789664414691E-001 | -1.925839668633162E-001 |
| 9 | 1.527248552219977E-000 | 3.969648696293301E-003 |
| 10 | 6.691341326208045E-000 | 2.097266548250023E-003 |

The potential (12) was examined within the frame of VM and the same bound energy of -5.4934228 MeV was received. The variational parameters and expansion coefficients of the radial wave function are listed in Table 4. The asymptotic constant at distances of 5-20 fm turned out to be 2.09(1) and the residual error is of the order of 10$^{-13}$.

Table 4
The variational parameters and expansion coefficients of the radial WF of the bound state of the p$^2$H system for the potential (12). The normalization of the function with these coefficients at the range 0-25 fm equals N=0.999999998

| *i* | $\beta_i$ | $C_i$ |
|---|---|---|
| 1 | 3.485070088054969E-001 | -1.178894628072507E-001 |
| 2 | 1.739943603152822E-002 | -6.168137382276252E-003 |
| 3 | 8.973931554450264E-003 | -4.319325351926516E-004 |
| 4 | 5.977571392609325E-002 | -7.078243409099880E-002 |
| 5 | 3.245586616581442E-002 | -2.743665993408441E-002 |
| 6 | 5.837991732045449E+001 | 1.102401456221556E-003 |
| 7 | 1.100441373510820E-001 | -1.384847981550261E-001 |
| 8 | 2.005318455817479E-001 | -2.114723533577409E-001 |

| 9 | 1.995655373133832E-000 | 3.955231655325594E-003 |
|---|---|---|
| 10 | 8.74651544040529E-000 | 2.101576342365150E-003 |

For the real bound energy in this potential it is possible to use the value -5.4934229(1) MeV with the calculation error of finding energy by two methods equal to ±0.1 eV, because the variational energy decreases as the dimension of the basis increases and gives the upper limit of the true bound energy, but the finite-difference energy increases as the size of steps decreases and the number of steps increases.

*3.2. Astrophysical S-factor*

In our calculations we considered the energy range of the radiative p$^2$H capture from 1 keV to 10 MeV and found the value of 0.165 eV b for the $S$(E1)-factor at 1 keV for the potentials from Table 1. The value found is slightly lower than the known data if we consider the total S-factor without splitting it into $S_s$ and $S_p$ parts resulting from M1 and E1 transitions. This splitting was made in work [22], where $S_s$(0)=0.109(10) eV b and $S_p$(0)=0.073(7) eV b, which gives the value of 0.182(17) eV b for the total S-factor. At the same time, the authors give the following values $S_0$=0.166(5) eV b and $S_1$=0.0071(4) eV b keV$^{-1}$ in the linear interpolation formula

$$S(E_{c.m.}) = S_0 + E_{c.m.} \cdot S_1 , \qquad (14)$$

and for $S$(0) leads to the value of 0.166(14) keV b, which was received taking into account all possible errors. The results with the splitting of the S-factor into $M$1 and $E$1 parts are given in one of the first of works [20], where $S_s$=0.12(3) eV b, $S_p$=0.127(13) eV b for the total $S$-factor 0.25(4) eV b.

As it can be seen, there is a visible difference between these results, so, in future we will take as a reference point the total value of S-factor at zero energy which was measured in various works. Furthermore, the new experimental data [23] lead to the value of total S(0)=0.216(10) eV b and this means that contributions of M1 and E1 will change. The following parameters of linear extrapolation (14) are given in this work $S_0$=0.216(6) eV b and $S_1$=0.0059(4) eV b keV$^{-1}$, that are noticeably differ from the data of work [22].

The known extractions of the S-factor from the experimental data, without splitting to M1 and E1 parts, at zero energy give the value of 0.165(14) eV b [31]. The previous measurements by the same authors lead to the value 0.121(12) eV b [32], and for theoretical calculations of work [33] the values $S_s$=0.105 eV b, $S_p$=0.08-0.0865 eV b are received for different models.

One can see that the experimental data over the last 10-15 years is very ambiguous. These results allow to come to a conclusion that, most probably, the value of total $S$-factor at zero energy is in the range 0.11-0.23 eV b. The average of these experimental measurements equals 0.17(6) eV b what is in a good agreement with the value calculated here on the basis of the $E$1 transition only.

Our calculation results for the $S$-factor of the p$^2$H radiative capture with the potential from table 1 at the energy range from 1 keV to 10 MeV are shown in Figs. 1 and 2 by dotted lines. Now the calculated S-factor reproduces experimental data at the energies down to 10-50 keV [22] comparatively well and at lower energies the calculated curve practically falls within the experimental error band of work [23].

Solid lines in Figs. 1 and 2 show the results for potential (12) which describes the behavior of the $S$-factor somewhat better at energies from 50 keV to 10 MeV and which gives the value of $S$=0.135 eV b for the energy of 1 keV. At energies of 20-50 keV the calculation curve follows the line of the lower limit of the error band of work [22], and at the energies below 10 keV it falls within the experimental error band of the LUNA project which was received recently [23]. The value of the S-factor at zero energy of this potential is in a good agreement with the data of the $S_p$ from work [20] for the $E1$ transition.

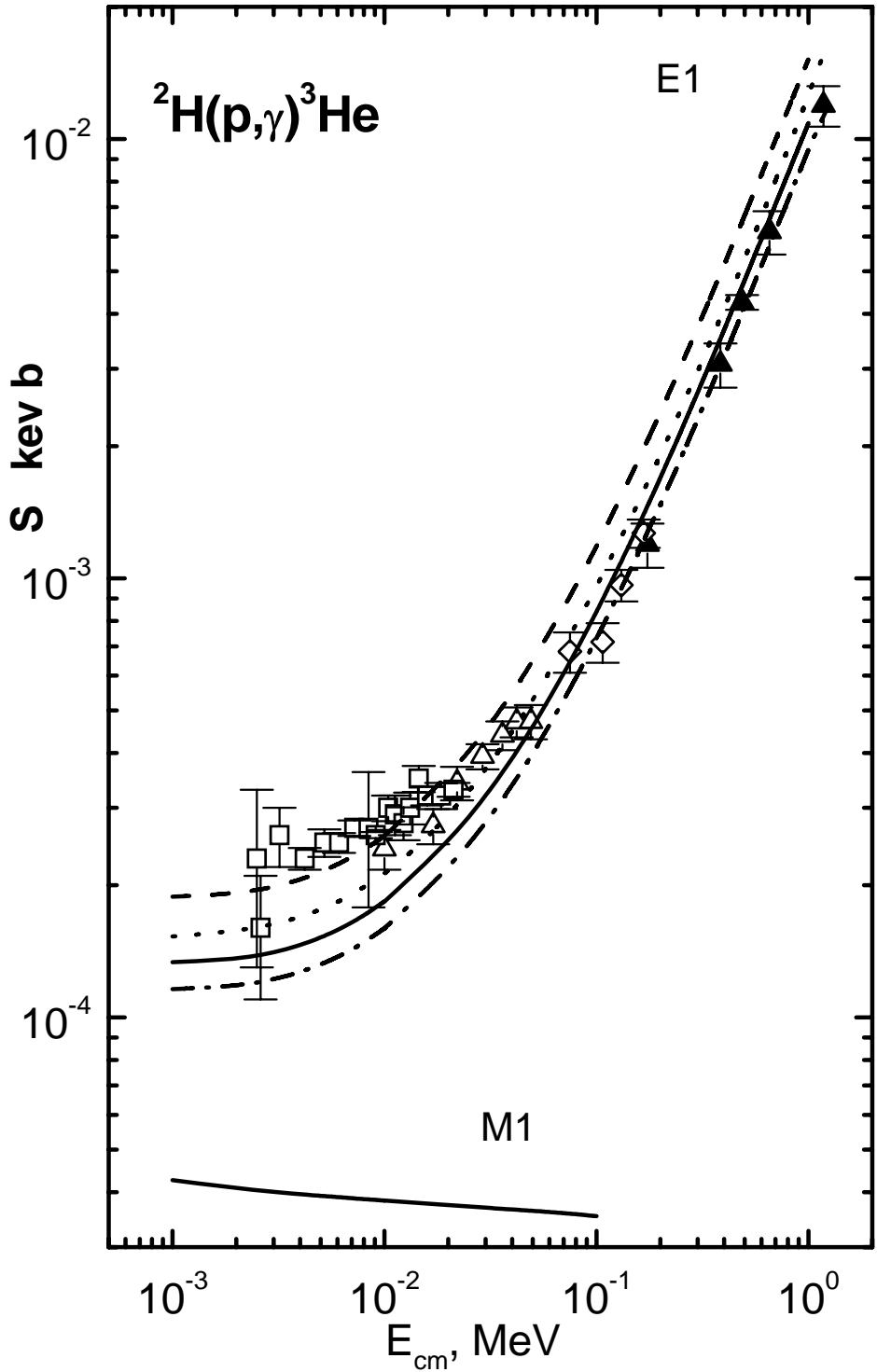

Fig. 1. Astrophysical S-factor of p²H radiative capture in the range 1 keV-1 MeV. Lines: calculations with the potentials mentioned in the text. Triangles denote the experimental data from [20], blank rhombs from [21], blank triangles from [22], blank blocks from [23].

The dashed lines in Figs. 1 and 2 show the results for potential (13) and the dash-dotted lines show those for potential (11). From these calculations one may conclude that the best results are obtained with the BS potential (12) which describes the experimental data in the widest energy range. It represents a sort of a compromise in describing asymptotic constant, charge radius and S-factor of the radiative p²H capture.

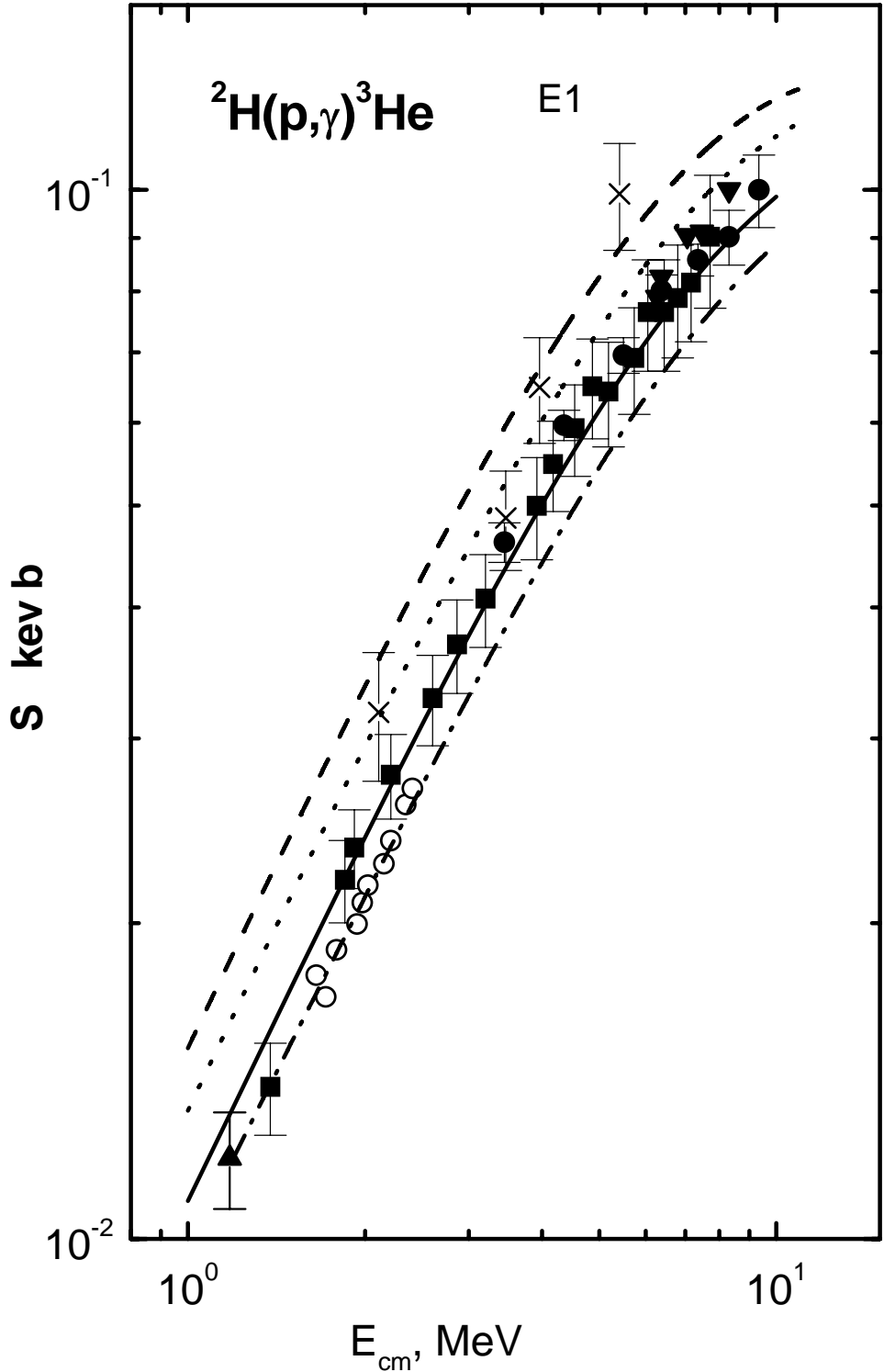

Fig. 2. Astrophysical S-factor of p$^2$H radiative capture in the range 1 MeV-10 MeV. Lines: calculations with the potentials mentioned in the text. Triangles denote the experimental data from [20], squares are from work [34], black points - from [35], crosses - from [36], inverted triangles - from [37], blank circles - from [38].

As it is seen in Fig. 1, the *S*-factor at the low energy range near 1-2 keV is practically constant and thus defines the *S*-factor value at zero energy which turns out to be approximately the same as its value at the energy equal to 1 keV. So, it seems that the difference between the values of the *S*-factor at energies 0 and 1 keV should be at most 0.05 eV b, and this value can be considered as the calculation error of the *S*-factor at zero energy.

The $M$1 transition from the $S$ scattering state, which is mixed in accordance with Young's schemes, to the bound state, which is "pure" according to orbital symmetries of the $S$ state of the $^3$He nucleus, can give a contribution to the astrophysical $S$-factor at low energies. For our calculations we used the doublet $S$-potential of the scattering states with the parameters listed in Table 1 and the BS potential (12). The calculation results at the energies 1-100 keV are shown in Fig. 1 by the solid line at the bottom of the figure. It can be seen that the cross-section of the $M$1 process is several times lower than the cross-section of the $E$1 transition.

However, it is necessary to note that we are unable to build the scattering S-potential uniquely because of the ambiguities in the results of different phase shift analyses. The other variant of potential with parameters $V_0$=-55.0 MeV and $\alpha$=0.2 fm$^{-2}$ [10,17], which also describes well the $S$ phase shift, leads at these energies to cross-sections of the $M$1 process several times higher than those of $E$1.

Such a big ambiguity in parameters of the $S$-potential of scattering, associated with errors of phase shifts extracted from the experimental data, does not allow us making certain conclusions about the contribution of the $M$1 process in the p$^2$H radiative capture. If the BS potentials are defined by the bound energy, asymptotic constant and charge radius quite uniquely and the potential description of the scattering phase shifts, which are "pure" in accordance with Young's schemes, is an additional criteria for determination of such parameters, then, for the construction of the scattering potential it is necessary to carry out a more accurate phase shift analysis for the $^2S$-wave and to take into account the spin-orbital splitting of $^2P$ phase shifts at low energies, as it was done for the elastic p$^{12}$C scattering at energies 0.2-1.2 MeV [39]. This will allow us to adjust the potential parameters used in the calculations of the p$^2$H capture in the potential cluster model, the results of the calculations of which depend strongly on the accuracy of the construction of the interaction potentials in accordance with the scattering phase shifts.

Thus, the $S$-factor calculations of the p$^2$H radiative capture for the $E$1 transition at the energy range down to 10 keV, which we carried out about 15 years ago [6], when the experimental data above 150-200 keV were only known, are in a good agreement with the new data of works [21, 22] in the energy range 10-150 keV. And this is true about both the potential from Table 1 and the interaction with parameters from (12). The results for the two considered potentials at the energies lower than 10 keV practically fall within the error band of work [23] and show that the $S$-factor tends to remain constant at energies 1-3 keV.

## 4. p³H radiative capture

Now let's consider the possibility of description of the astrophysical S-factor of p³H radiative capture at the energy range down to 1 keV as a continuation of the theoretical investigation of thermonuclear reactions [17] on the basis of the potential cluster model with the splitting of orbital states according to Young's schemes [41]. This reaction can be of some interest for a good understanding of the nature of thermonuclear photoprocesses with lightest atomic nuclei at low energies, both from theoretical and experimental points of view. Thus, experimental researches into this reaction are continued, and quite recently the new data for the total cross-sections of the p³H radiative capture and for the astrophysical S-factor at the energy range down to 12 keV (c.m.) were obtained.

### 4.1. Potentials and phase shifts of scattering

To calculate photonuclear processes in the systems under consideration the nuclear part of the potential of inter-cluster p³H and p³He interactions is represented as (5) with a point-like Coulomb potential. The potential of each partial wave, as for the previously considered p²H system, was constructed so as to describe correctly the respective partial phase shift of the elastic scattering [40].

As a result, the potentials of the p³He interaction for scattering processes which are "pure" in accordance with $T=1$ were received. The parameters of such potentials are fully listed in Table 5 [41, 42]. The singlet $S$ phase shift of elastic p³He scattering which is "pure" in accordance with isospin is shown in Fig. 3 by the solid line together with the experimental data of works [43-45]. Further it is used for receiving the singlet p³H phase shifts which are "pure" in accordance with $T=0$.

Table 5
The singlet potentials of the p³He system which are "pure" in accordance with isospin T=1 [42]

| System | $^{2S+1}L$ | $V_0$ (MeV) | $\alpha$ (fm$^{-2}$) | $V_1$ (MeV) | $\gamma$ (fm$^{-1}$) |
|---|---|---|---|---|---|
| P³He | $^1S$ | -110.0 | 0.37 | +45.0 | 0.67 |
|  | $^1P$ | -15.0 | 0.1 | --- | --- |

Since there are several variants of the phase shift analyses [40, 43-45] for the $^1P_1$ singlet waves, the parameters of potentials from Table 5 are chosen so as to lead to a kind of a compromise between the different phase shift analyses. The singlet $^1P_1$ phase shift of elastic p³He scattering with $T=1$, used in our calculations of the $E1$ transition to the ground state (GS) of the ⁴He nucleus in the p³H channel with $T=0$, is shown in Fig. 4 by the solid line together with the experimental data of works [43-48].

Because of the fact that the p³H system is isospin-mixed the singlet and triplet phase shifts, and consequently the potentials, effectively depend on two values of isospin. The result of mixing in terms of isospin is mixing in terms of Young's schemes. In particular, two orbital schemes {31} and {4} are allowed in the singlet state.

The isospin-mixed singlet $S$ phase shift of the elastic p³H scattering which was calculated from the experimental differential cross-sections and which we used for receiving "pure" p³H phase shifts is shown in Fig. 5 by the solid line with the experimental data from works [49-51]. We use the first set of phase shifts of scattering from work [51]. The following parameters were received: $V_0 = -50.0$ MeV, $\alpha = 0.2$ fm$^{-2}$.

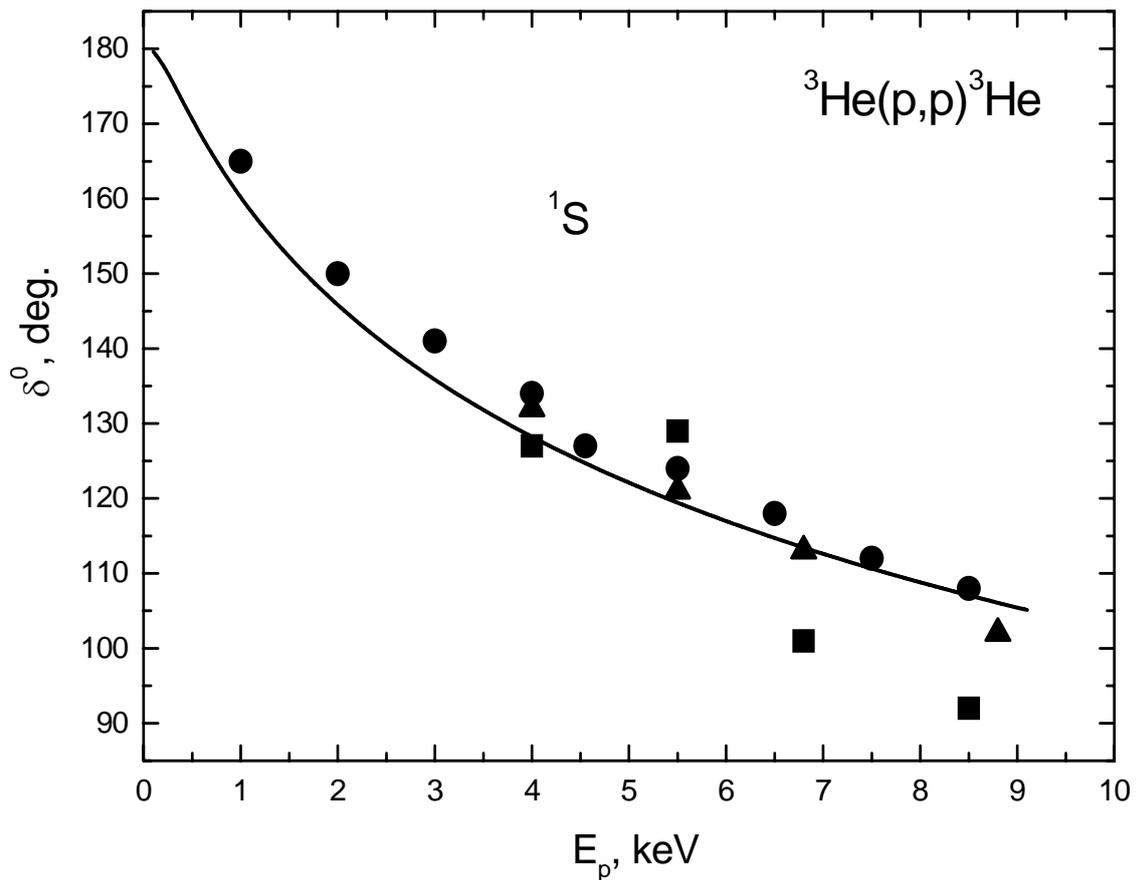

Fig. 3. The singlet $^1S$ phase shift of the elastic p$^3$He scattering. Black points denote the experimental data from [43], blocks are from work [44], triangles – from work [45].

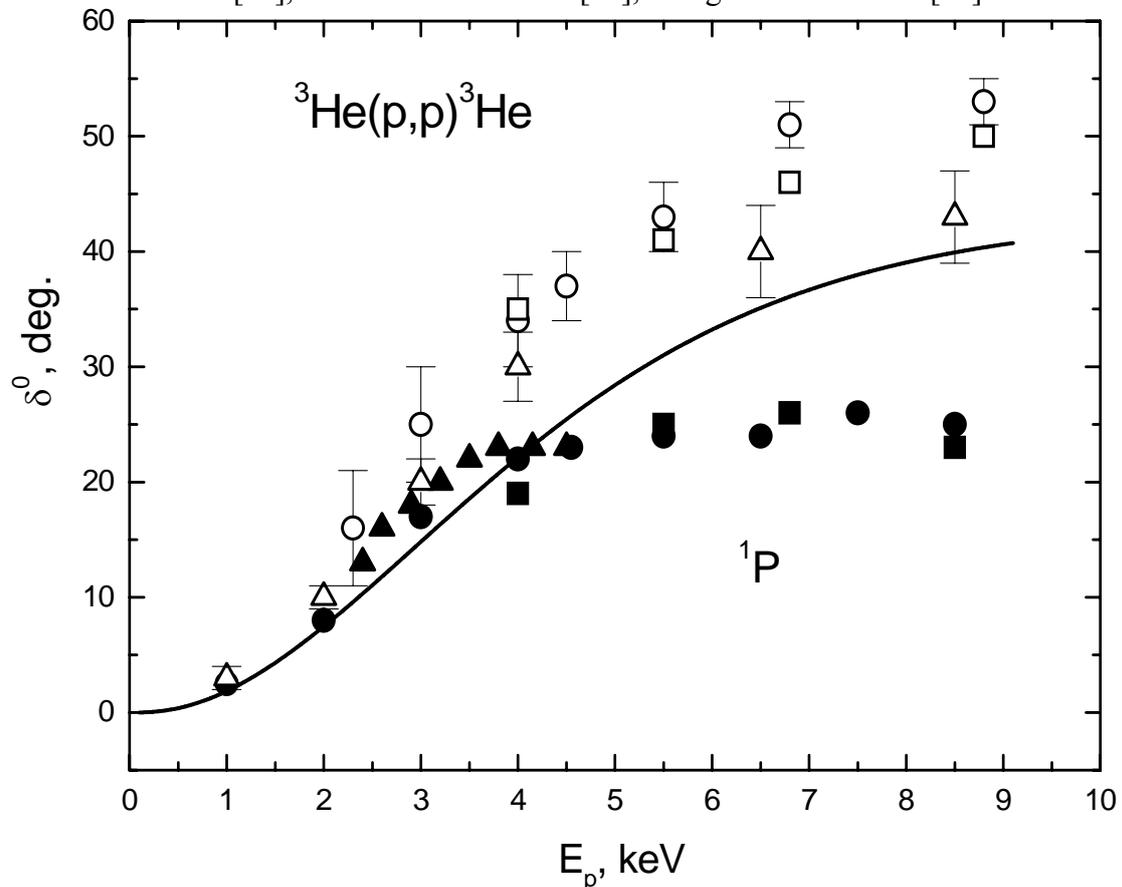

Fig. 4. The singlet $^1P$ phase shift of the elastic p$^3$He scattering. Black points denote the experimental data from [43], blocks are from work [44], triangles - from [46], blank circles - from [47], blank blocks - from [45], blank triangles - from [48].

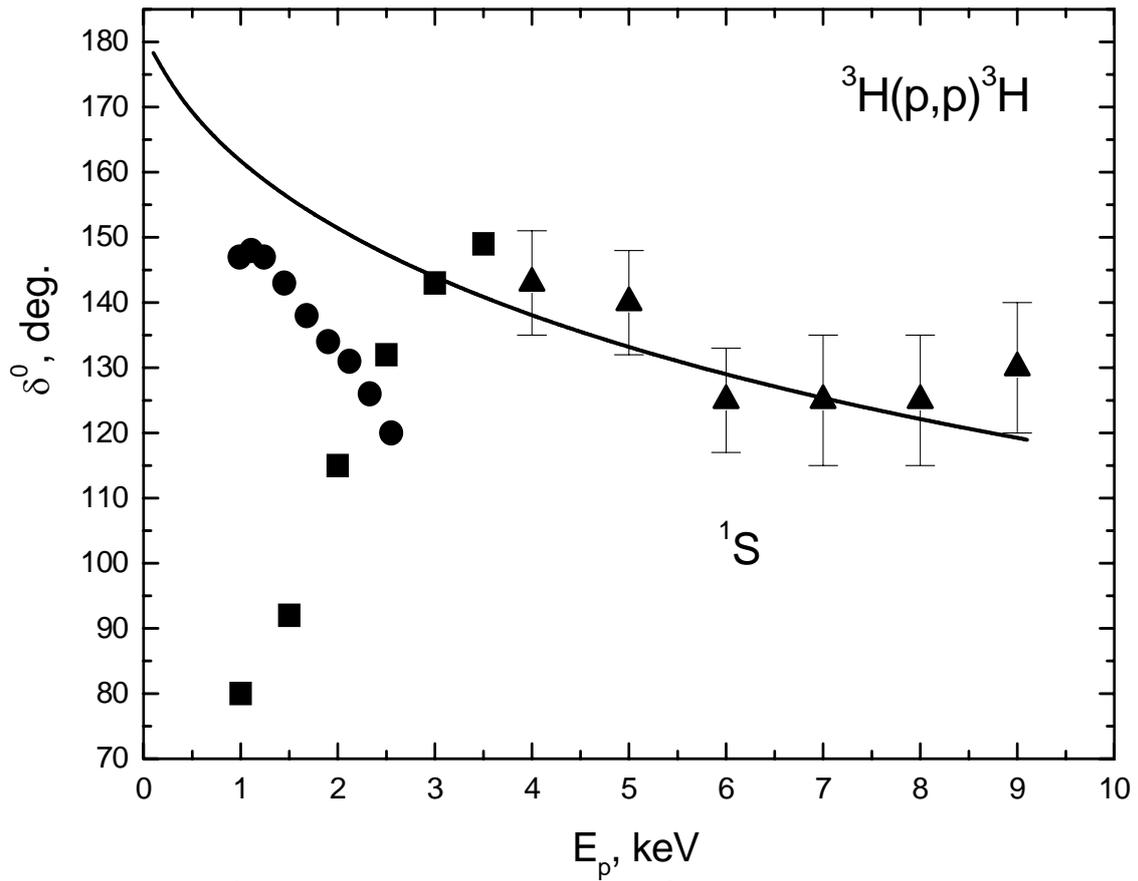

Fig. 5. The singlet $^1S$ phase shift of the elastic p$^3$H scattering. Black points denote the experimental data from [49], blocks - from work [50], triangles - from work [51].

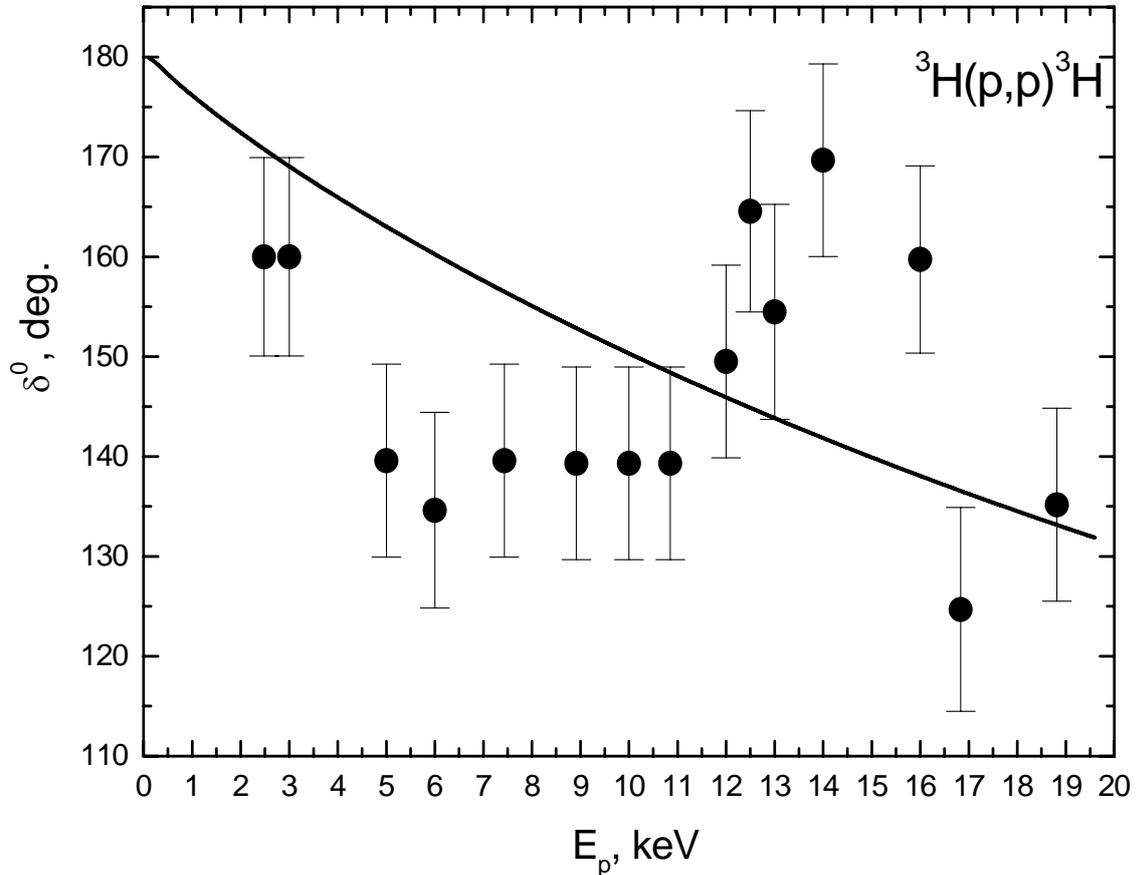

Fig. 6. The singlet $^1S$ phase shift of the elastic p$^3$H scattering "pure" in terms of Young's schemes.

As it was shown in works [10,42], the isospin-mixed singlet phase shifts of the p$^3$H scattering can be represented as a half-sum of isospin-pure singlet phase shifts

$$\delta^{\{T=1\}+\{T=0\}} = 1/2\delta^{\{T=1\}} + 1/2\delta^{\{T=0\}}, \quad (15)$$

which is equivalent to the expression

$$\delta^{\{4\}+\{31\}} = 1/2\delta^{\{31\}} + 1/2\delta^{\{4\}}.$$

It is usually considered [41] that the "pure" phase shifts correspond to the Young's schemes {31} with $T=1$ for the p$^3$He system and {4} with $T=0$ for the p$^3$H system. The isospin-pure phase shifts of p$^3$H scattering with $T=0$ are constructed on the basis of expression (15) by using known "pure" phase shifts of scattering with $T=1$ for the p$^3$He system [43-45] and "mixed" p$^3$H phase shifts with $T=0$ and 1 [49-51]. They are in turn used for obtaining "pure" p$^3$H interaction potentials [42]. In particular, for the $^1S$ wave the following parameters were received:

$$V_0 = -63.1 \text{ MeV}, \alpha = 0.17 \text{ fm}^{-2}, V_1 = 0. \quad (16)$$

In Fig. 6 the singlet $^1S$ phase shift of the elastic p$^3$H scattering with "pure" Young's schemes is shown with the black points, and its calculation results with the potential (16) are shown with the solid line. Such "pure" interactions can be used for the calculations of various characteristics of the bound state of the $^4$He nucleus in the p$^3$H channel and the results will depend on the clusterization rate of this nucleus into the considered channel.

The interactions received in [10, 42] generally give a correct description of the channel bound energy of the p$^3$H system (with an accuracy about few keV) and the root-mean-square radius of the $^4$He nucleus [42]. The calculations of the differential [41] and total cross-sections of the radiative p$^3$H capture and astrophysical $S$-factors were done with these potentials at the energy range down to 10 keV [42]. Though, at that time we only knew $S$-factor experimental data in the range above 700 keV [52].

A short time ago the new experimental data in the energy range from 50 keV to 5 MeV [53] and at 12 and 36 keV [54] appeared. That is why, it was interesting to know if it is possible to describe the new data on the basis of the potential cluster model, with the earlier obtained singlet $^1P$ potential and adjusted interaction of the ground $^1S$ state.

Our preliminary results show that for calculations of the $S$-factor at energies of the order of 1 keV we have to meet the same requirements as for the p$^2$H system (which were discussed in the previous section) and first of all - to raise an accuracy of finding the $^4$He bound energy in the p$^3$H channel. So, by using the new modified computer programs, we adjusted the parameters of the ground state potential of the p$^3$H system in the $^4$He nucleus. These potentials differ from the potentials in work [42] by 0.2 MeV and are listed in Table 6.

Basically, this difference results from the use of the exact mass values of proton and $^3$H particles [12] in new calculations and more accurate description of the bound state energy of the $^4$He nucleus. The value -19.813810 MeV was obtained experimentally for the bound state of the $^4$He nucleus in the p$^3$H channel on the basis of exact mass values [12], and the calculation with such a potential leads to the value of -19.81381000 MeV. The absolute accuracy of searching for the bound energy in our computer program based on the finite-difference method was taken to be at the level of $10^{-8}$ MeV.

Table 6
The isospin-pure potentials of the p$^3$H [42] interactions in the singlet channel with T=0. $E_{BS}$ is the calculated energy of the bound state, $E_{EXP}$ - its experimental value [25], the depth of attractive part of the potential (5) $V_1$=0.

| System | $L$ | $V_0$ (MeV) | $\alpha$ (fm$^{-2}$) | $E_{BS}$ (MeV) | $E_{EXP}$ (MeV) |
|---|---|---|---|---|---|
| p$^3$H | $^1S$ | -62.906841138 | 0.17 | -19.81381000 | -19.813810 |
|  | $^1P$ | +8.0 | 0.03 | --- | --- |

The calculation accuracy of the "tail" of the WF of BS of the p$^3$H system was verified using asymptotic constant $C_W$ with Whittaker asymptotics (8) [13, 55], and its value in the range of 5-10 fm turned out to be $C_W$=4.52(1). The experimental data known for this constant in the p$^3$H channel give the value of 5.16(13) [13]. In the same work for the asymptotic constant of the n$^3$He system there was obtained almost the same value 5.1(4).

At the same time, the asymptotic constant of the n$^3$He system in work [55] equals $C_W$=4.1 and if we suppose that the constants of p$^3$H and n$^3$He channels have a small difference, as it is shown in [13], then the value of the asymptotic constant of the p$^3$H channel should be in the range of 4.2 - 4.4, what is in a good agreement with our results. As it is seen, there is a big difference between the experimental results of asymptotic constants. For the n$^3$He system the asymptotic constant is in the range of 4.1-5.5, and for the p$^3$H channel it seems to be in the range from 4.2-4.4 to 5.3. For the $^4$He charge radius the value of 1.73 fm was obtained with the radius of tritium being 1.63 fm [24] and that of proton - 0.877 fm [12], while the experimental value of the $^4$He charge radius is 1.671(14) fm [25] (see Table 2).

The variational method with the expansion of the WF of the relative cluster motion in nonorthogonal Gaussian basis is used for an additional control of the accuracy of bound energy calculations for the *S*-potential from table 6, which allowed obtaining the bound energy of -19.81380998 MeV by using independent variation of parameters and the grid with dimension 10. The asymptotic constant $C_W$ (8) of the variational WF at distances of 5-10 fm remains at the level of 4.52(2). The variational parameters and expansion coefficients of the radial wave function having form (7) are listed in table 7.

Table 7
The variational parameters and expansion factors of the radial WF of the bound state of the p$^3$H system for the *S*-potential from Table 6. The normalization of the function with these factors in the range 0-25 fm equals N=0.9999999998

| $i$ | $\beta_i$ | $C_i$ |
|---|---|---|
| 1 | 3.775399682294165E-002 | -3.553662130779118E-003 |
| 2 | 7.390305111120065E-002 | -4.689092850709087E-002 |
| 3 | 1.377393687979590E-001 | -1.893147614352133E-001 |
| 4 | 2.427238748079469E-001 | -3.619752356073335E-001 |
| 5 | 4.021993911220914E-001 | -1.988757841748206E-001 |
| 6 | 1.780153251456691E+000 | 5.556224701527299E-003 |
| 7 | 5.459871888661887E+000 | 3.092889292994009E-003 |
| 8 | 1.921317723809205E+001 | 1.819890982631486E-003 |
| 9 | 8.416117121198026E+001 | 1.040709526875803E-003 |
| 10 | 5.603939880318445E+002 | 5.559240350868498E-004 |

For the real bound energy in this potential it is possible to use the value -19.81380999(1)

MeV with the calculation error of finding energy by the two methods used equals to ±0.01 eV, because, as we mentioned in the previous section, the variational energy decreases as the dimension of the basis increases and gives the upper limit of the true bound energy, but the finite-difference energy increases as the size of steps decreases and the number of steps increases.

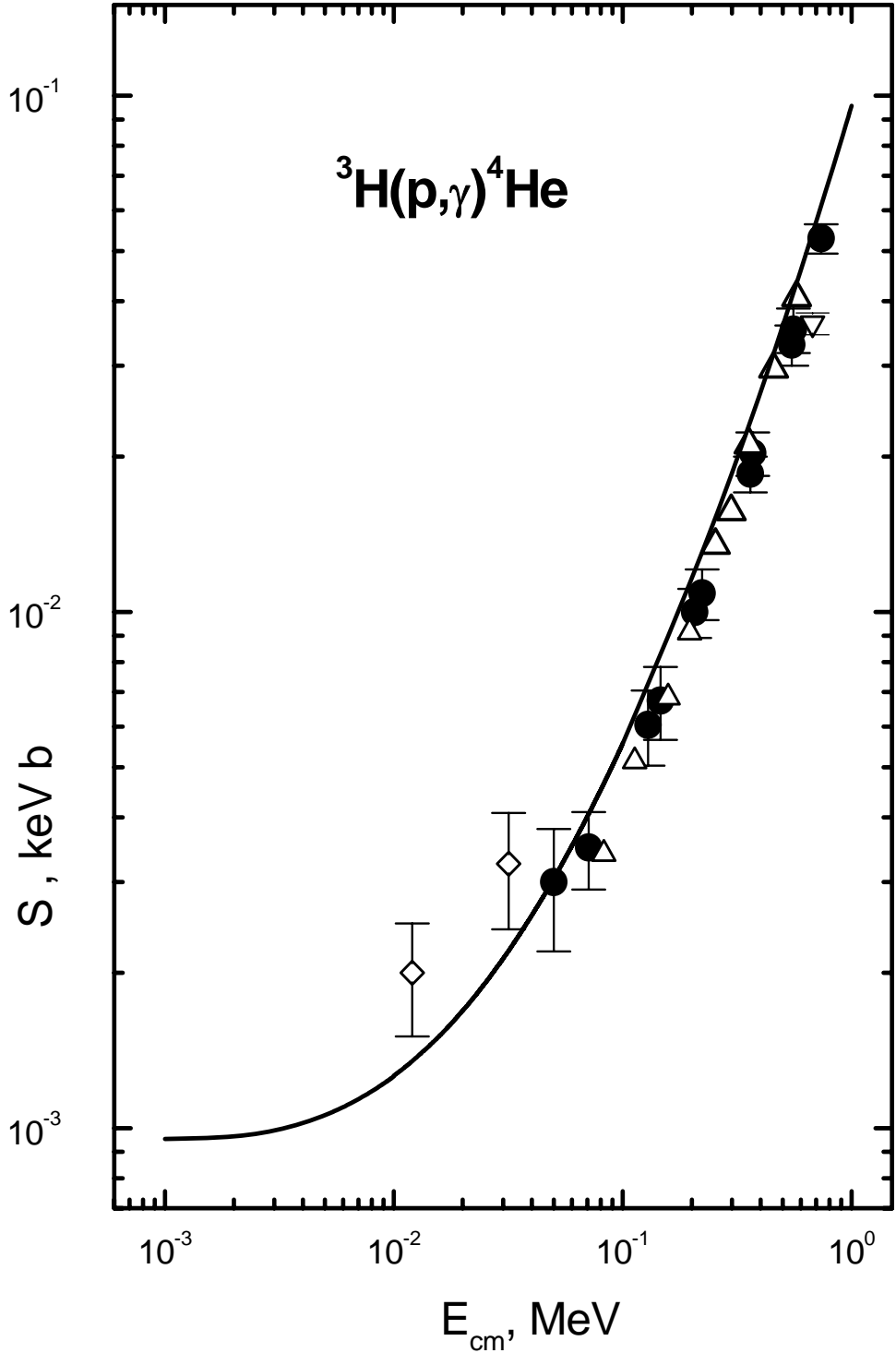

Fig. 7. Astrophysical S-factor of p$^3$H radiative capture in the range 1 keV – 1 MeV. Line: calculation with the potential mentioned in the text. Black dots represent the conversion of the total capture cross-sections from work [53] given in work [54], blank triangles - from [57], blank rhombs - from [54], inverted blank triangle - from [52].

It can be seen from the given results that the simple two-cluster p$^3$H model with the classification of orbital states according to Young's schemes allows obtaining quite reasonable values for such characteristics of the bound state of the $^4$He nucleus as charge radii and asymptotic constants. Thus, these results are indicative of a comparatively high clusterization rate of the nucleus into the p$^3$H channel.

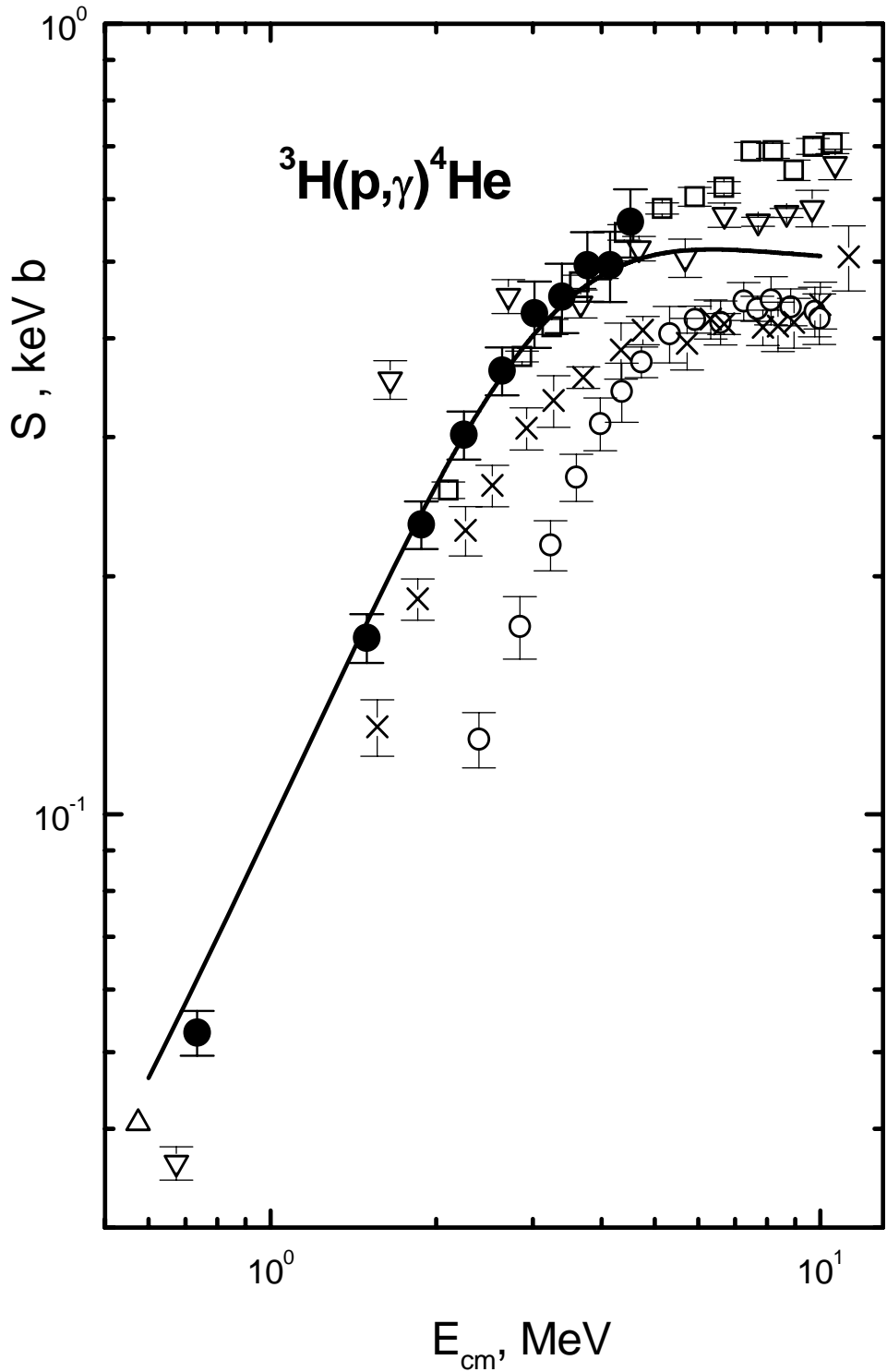

Fig. 8. Astrophysical $S$-factor of p$^3$H radiative capture in the range 1 MeV - 10 MeV. Black dots is the conversion of the total capture cross-sections from work [53] given in work [54], blank triangles - from [57], blank circles - from [58], blank blocks - from [59], crosses (×) from [60], inverted blank triangles - from [52].

*4.2. Astrophysical S-factor*

Earlier, the total cross-sections and astrophysical *S*-factor of the process of p$^3$H radiative capture were considered on the basis of the potential cluster model in work [42]. It was assumed that the transitions with isospin changes $\Delta T=1$ [56] give the main contribution to the *E*1 cross-sections of the $^4$He photodecay into the p$^3$H channel or to the radiative p$^3$H capture. Thus, we should use the $^1P_1$ scattering potential of the isospin-pure singlet state of the p$^3$He system with *T*=1 and the $^1S$ potential of the isospin-pure ground state of the $^4$He nucleus in the p$^3$H channel with *T*=0 [42].

Using this concept we have carried out the calculations of the *E*1 transition with the adjusted potential of the $^4$He ground state from Table 6. The calculation results for the astrophysical S-factor at the energies down to 1 keV are shown in Figs. 7 and 8 by solid lines. These results are almost the same as the previous ones which we obtained in work [42] at the energy range down to 10 keV. The new experimental data are taken from works [53, 54] and the data which we used additionally are from [57].

As one can see from the figures, the calculation results, which we obtained about 15 years ago, reproduced the new *S*-factor data obtained in work [53] at the energy range from 50 keV to 5 MeV (c.m.) very well. The ambiguity of these data is visibly less at the energy range over 1 MeV when compared with the previous results of works [52, 58-60], and they give a more accurate overall behavior of the *S*-factor at low energies, practically coinciding with the previous data [57] at the energy range 80 - 600 keV. The energy range over 1-2 MeV was measured in many works, so, in Fig. 8, we give these results for comparison: blank circles are from [58], blank blocks - from [59], inverted blank triangles - from [52], crosses (×) from [60].

At the energy of 1 keV the value of the *S*-factor turned out to be equal to 0.95 eV b, and the calculation results of its value at the energy range lower than 50 keV lie slightly below the new data [54], where for the *S*(0) the value of 2.0(2) eV b was received. Note that a simple extrapolation of existing experimental data to 1 keV of the last three points of works [53,57] leads to its value of about 0.6(3) eV b, which is three times less than the results of work [54].

As it is seen in Fig. 7 the *S*-factor at the lowest energies, approximately in the range of 1-3 keV, is practically constant. Thus, the *S*-factor value at zero energy can be defined almost the same as its value at the energy of 1 keV. As a result, the difference between the values of the *S*-factor at energies 0 and 1 keV seems to be at most 0.05 eV b, and this value can be considered as the calculation error of the *S*-factor at zero energy.

Thus, on the basis of only *E*1 process, we have managed to predict the general behavior of the S-factor of the p$^3$H capture at the energy range from 50 keV to 700 keV, because our calculations in the energy range down to 10 keV were done about 15 years ago [42], when only experimental data higher than 700 keV were known. The results of the calculations at the energy range from 50 keV to 5 MeV are in a good agreement with the new data of the *S*-factor from work [53] (black dots in Figs. 7, 8).

# 5. p⁶Li radiative capture

For the updating of the existing experimental data the new measurements of differential cross-sections of the elastic p$^6$Li scattering at the energy range from 350 keV to 1.15 MeV (l.s.) with 10% errors were done. The experimental data [3], which we are going to consider in this section, are received at five energy values: 593 keV for 13 angles of scattering in the range $57^0$-$172^0$, 746.7 and 866.8 keV for 11 angles in the range $45^0$-$170^0$ and at energies 976.5 and 1136.6 keV for 15 angles in the range $30^0$-$170^0$.

On the basis of measurements from work [3] and differential cross-sections of the elastic scattering at the energy 500 keV from earlier work [61] we have carried out the phase shift analysis and received $^{2,4}S$ and $^2P$-phase shifts of scattering. The p$^6$Li interaction potentials for $L$=0 and 1 at low energies and without taking into account the spin-orbital splitting were constructed according to the phase shifts obtained, and then the calculations of the astrophysical $S$-factor at the energy above 10 keV were made.

## 5.1. Phase shift analysis

For scattering processes in the particle system with a spin 1/2+1, without taking into account the spin-orbital splitting, the cross-section of the elastic scattering is represented as [62]

$$\frac{d\sigma(\theta)}{d\Omega} = \frac{2}{6}\frac{d\sigma_d(\theta)}{d\Omega} + \frac{4}{6}\frac{d\sigma_q(\theta)}{d\Omega}, \quad (17)$$

were $d$ and $q$ are related to the doublet (with total spin 1/2) and quartet (with total spin 3/2) states of the p$^6$Li scattering and

$$\frac{d\sigma_d(\theta)}{d\Omega} = |f_d(\theta)|^2, \quad \frac{d\sigma_q(\theta)}{d\Omega} = |f_q(\theta)|^2. \quad (18)$$

The scattering amplitudes are written as

$$f_{d,q}(\theta) = f_c(\theta) + f_{d,q}^N(\theta), \quad (19)$$

where

$$f_c(\theta) = -\left(\frac{\eta}{2k\sin^2(\theta/2)}\right)\exp\{i\eta\ln[\sin^{-2}(\theta/2)] + 2i\sigma_0\},$$

$$f_d^N(\theta) = \frac{1}{2ik}\sum_L(2L+1)\exp(2i\sigma_L)[S_L^d - 1]P_L(\cos\theta),$$

$$f_q^N(\theta) = \frac{1}{2ik}\sum_L(2L+1)\exp(2i\sigma_L)[S_L^q - 1]P_L(\cos\theta), \quad (20)$$

and $S_L^{d,q} = \eta_L^{d,q}\exp(2i\delta_L^{d,q}(k))$ is the scattering matrix in the doublet or quartet spin state [62].

It is possible to use simple formulae (17-19) for the calculations of cross-sections of the elastic scattering because the spin-orbital phase shift splitting at low energies is quite insignificant. It is confirmed by the results of the phase shift analysis given in work [63] where the authors take into account the spin-orbital splitting of the scattering phase shifts.

Earlier the phase shift analysis of the differential cross-sections and the excitation functions of the elastic p$^6$Li scattering was made in work [63], but this analysis did not include the doublet $^2P$-wave. Our phase shift analysis is based on the differential cross-sections given only in works [3] and [61]. The calculations are made for lower energies having the importance for the nuclear astrophysics and take into account all partial waves, including the doublet $^2P$-wave.

The first energy we considered is 500 keV from work [61] and it leads to $^2S$ and $^4S$-phase shifts of scattering which are listed in Table 8 and quite reasonably describe the experimental results with small average value of $\chi^2$=0.15. The effort to take into account the doublet $^2P$ and quartet $^4P$-phase shifts leads to the low values of these shifts. The error of the differential cross-sections of these data was taken to be 10%.

Table 8
Results of the phase shift analysis of the elastic p$^6$Li scattering

| № | E, keV | $^2S$, deg. | $^4S$, deg. | $^2P$, deg. | $^4P$, deg. | $\chi^2$ |
|---|---|---|---|---|---|---|
| 1 | 500 | 176.2 | 178.7 | --- | --- | 0.15 |
| 2 | 593 | 174.2 | 178.8 | --- | --- | 0.15 |
| 3 | 746.4 | 170.1 | 180.0 | --- | --- | 0.23 |
|   | 746.4 | 172.5 | 179.9 | 1.7 | 0.0 | 0.16 |
| 4 | 866.8 | 157.8 | 180.0 | --- | --- | 0.39 |
|   | 866.8 | 170.2 | 174.9 | 3.9 | 0.0 | 0.22 |
|   | 866.8 | 169.6 | 175.0 | 3.5 | 0.1 | 0.23 |
| 5 | 976.5 | 160.0 | 178.5 | --- | --- | 0.12 |
|   | 976.5 | 167.0 | 174.5 | 1.1 | 0.0 | 0.12 |
|   | 976.5 | 166.9 | 174.5 | 1.1 | 0.0 | 0.12 |
| 6 | 1136.3 | 144.9 | 180.0 | --- | --- | 0.58 |
|   | 1136.3 | 164.7 | 171.1 | 5.8 | 0.0 | 0.32 |
|   | 1136.3 | 166.4 | 169.9 | 5.5 | 0.1 | 0.32 |

The next five energies are the new results of measurements of the differential cross-sections taken in work [3]. The first of them is equal to 593 keV and leads to the $^{2,4}S$-phase shifts which differ slightly from those for the previous energy, they have the same value of $\chi^2$ and are listed in Table 8. The phase shifts of $^{2,4}P$-waves vanish to zero.

The second energy 746.7 keV leads to the $^{2,4}S$-phase shifts (see Table 8) which allow us to describe the cross-sections with $\chi^2$=0.23. In spite of the small value of $\chi^2$, the attempt to take into account $^{2,4}P$-phase shifts was made. At the beginning we supposed that the quartet $^4P$-phase shift is negligible as it followed from the results of work [63] where their account begins from 1.0-1.5 MeV only. The results of our analysis taking into account the $^2P$-phase shift only are shown in Fig. 9 and are listed in Table 8. It can be seen that the small doublet $^2P$-phase shift slightly changes the doublet $^2S$-phase shift increasing its value and reducing the value of $\chi^2$ to 0.16. The effort to take into account the $^4P$-phase shift too leads to the negligible values (less than 0.1$^0$). This fact is absolutely in conformity with the results of work [63] and it will be demonstrated on the example of the next energy of 866.8 keV.

The results of the phase shift analysis at the energy of 866.8 keV taking into account the $^{2,4}S$-waves only are given in Table 8 at $\chi^2$=0.39. They suggest that the value of the $^2S$-phase shift falls sharply in comparison with the previous energy. However, if $^2P$-wave is taken into account, its value increases noticeably (Fig. 10 and Table 8) while the $\chi^2$ value decreases almost twofold. The effort to take into account the quartet $^4P$-phase shift leads to the value not more than $0.1^0$ (Table 8) which means that its contribution at this energy range is very small. Any change in this phase shift resulting in its increase leads to the increase in $\chi^2$, even at the different values of other phase shifts. For this energy and for all considered energies from work [3] it is impossible to find some non-zero doublet and quartet phase shifts whith the value of $\chi^2$ approaching its minimum.

The next considered energy equals 976.5 keV and, if $^{2,4}P$-waves are not taken into account, it leads to the values of $^2S$ and $^4S$-phase shifts listed in Table 8. The further consideration of the $^2P$-wave noticeably increases the value of the $^2S$-phase shift when the $^4P$-wave equals zero, as it seen in Fig. 11 and Table 8 at $\chi^2=0.12$. If we include the quartet $^4P$-wave in the analysis, then it also vanishes to zero as $\chi^2$ decreases.

The last energy from work [3] is equal to 1.1363 MeV and it leads to a comparatively small value of $\chi^2 = 0.58$ even if we take into account only $^{2,4}S$-waves in the analysis (see Table 8). However the account of the $^2P$-wave noticeably decreases this value, and the calculation results for the differential cross-sections are shown in Fig. 12 and are listed in Table 8. The attempt to take into account the $^4P$-phase shift leads to its negligible values in this case as well (see Table 8).

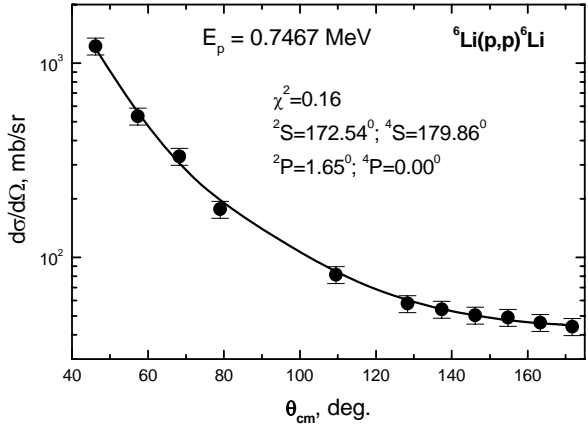
Fig. 9. Differential cross-sections of the elastic p$^6$Li scattering at energy 746.7 keV.

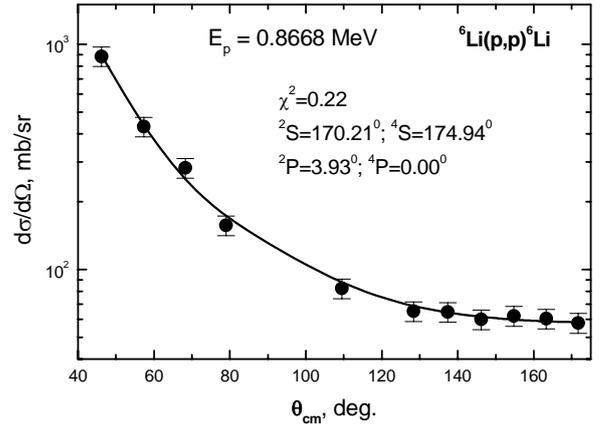
Fig. 10. Differential cross-sections of the elastic p$^6$Li scattering at energy 866.8 keV.

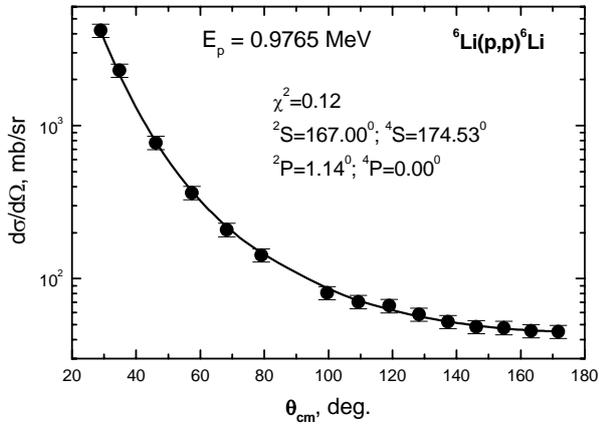
ig. 11. Differential cross-sections of the elastic p$^6$Li scattering at energy 976.5 keV.

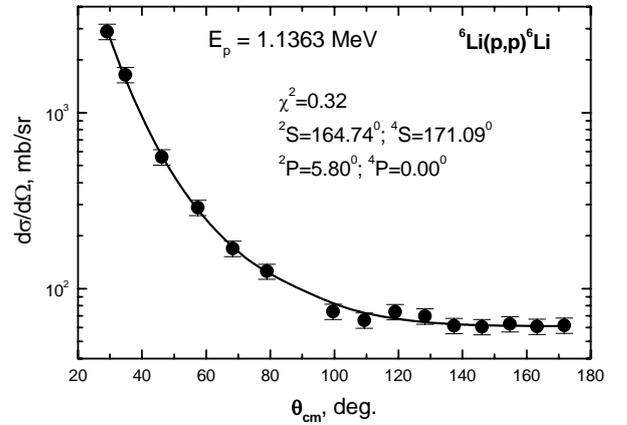
Fig. 12. Differential cross-sections of the elastic p$^6$Li scattering at energy 1136.3 keV.

Thus, al the experimental data of work [3] have a structure which does not require the presence of the quartet $^4P$-waves in this energy range, i.e. their value equals or lower than $0.1^0$.

This fact, in general, is in agreement with the results of work [63], but the doublet $^2P$-phase shift almost comes up to $6^0$ and it is impossible to neglect this phase shift, as it was done in [63].

The general pattern of the $^2S$ and $^4S$-phase shifts of scattering is shown in Fig. 13 and the doublet $^2P$-phase shifts are shown in Fig. 14. In spite of the large data spread for the $^4S$-phase shifts, the doublet $^2S$-phase shift tends to decrease, but significantly slower than it could be expected from the results of analysis [63], where the $^2P$-wave was not taken into account. If we do not take into account the doublet $^2P$-wave in our analysis, then the values of the $^2S$-phase shift are very close to the results of the phase shift analysis of work [63].

Errors of the phase shift analysis which are shown in Fig. 13 are due to the ambiguity of the phase shift analysis - it is possible to obtain slightly different values of the phase shifts of scattering with approximately the same value of $\chi^2$. This ambiguity is estimated as $1^0$-$1.5^0$ and is shown for the $S$ and $^2P$-phase shifts in Figs. 13 and 14.

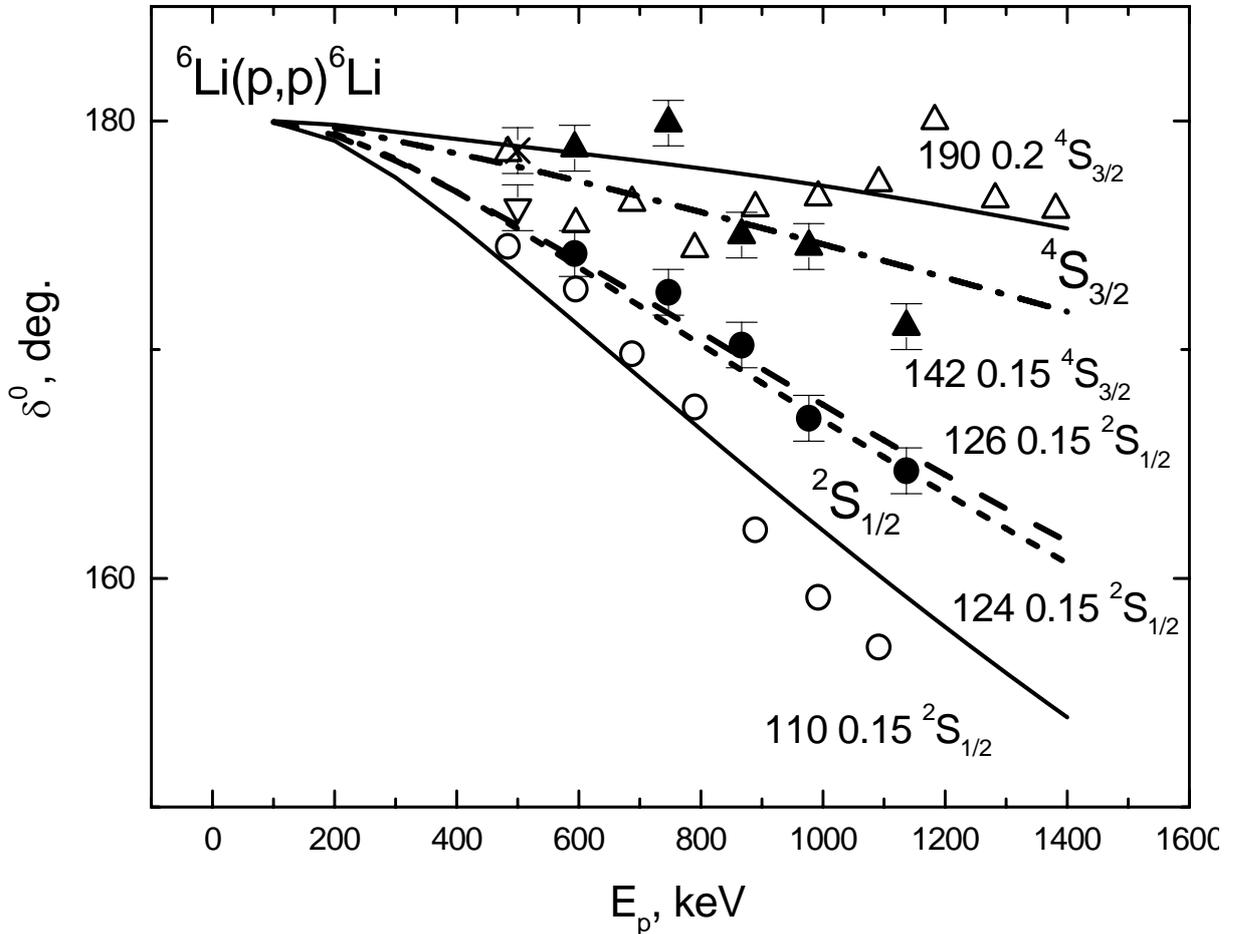

Fig. 13. Doublet and quartet $S$-phase shifts of the elastic p$^6$Li scattering at low energies. Doublet and quartet $S$-phase shifts taking into account $^2P$-wave, when $^4P$-phase shift was taken to be zero, are shown in the figure. Cross ($^4S$) and inverted blank triangle ($^2S$-phase shifts) are from data of work [61], black points ($^2S$) and triangles ($^4S$) are from [3]. For comparison the results of phase shift analysis [63] are represented by blank triangles and blank circles. Lines: calculation results for different potentials.

*5.2. Potential description of the phase shifts of scattering*

To calculate the partial inter-cluster p$^6$Li interactions according to the existing phase shifts of scattering we use common Gaussian potential with a point-like Coulomb component, which can be represented as (6). The following parameters for the description of the results of the phase shift analysis of work [63] were received:

$^2S$ - $V_0$=-110 MeV, $\alpha$=0.15 fm$^{-2}$,
$^4S$ - $V_0$=-190 MeV, $\alpha$=0.2 fm$^{-2}$.

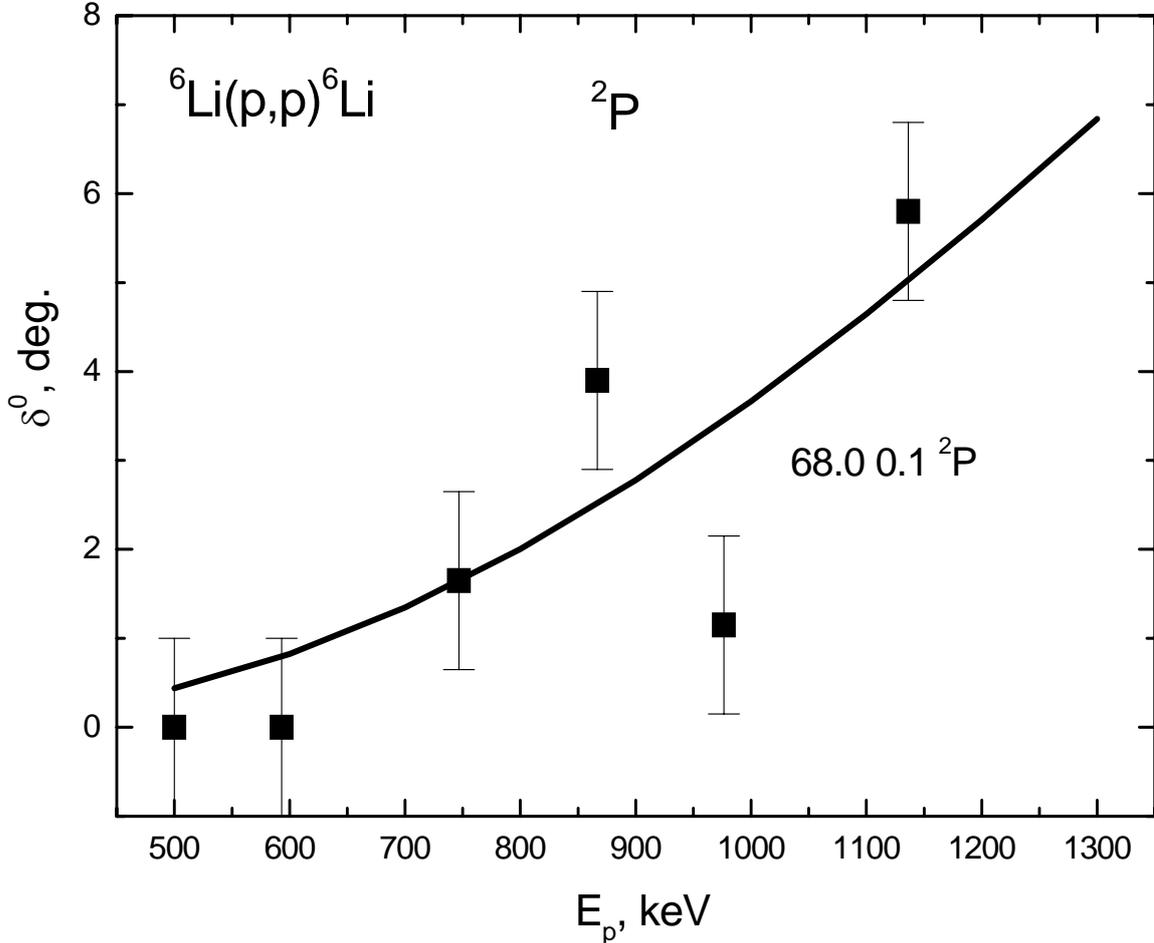

Fig. 14. Doublet $^2P$-phase shifts of the elastic p$^6$Li scattering at low energies. Blocks - results of our phase shift analysis at $^4P$=0. Lines - the result of calculation with received potentials.

They includ two forbidden bound states which correspond to the Young's schemes {52} and {7} [10,64]. The calculation results of the phase shifts for these potentials are shown in Fig. 13 by the solid lines with the results of the phase shift analysis [63] shown by blank circles and blank triangles.

For the description of our results of the phase shift of scattering the next potential parameters are preferable:

$^2S$ - $V_0$=-126 MeV, $\alpha$=0.15 fm$^{-2}$
$^4S$ - $V_0$=-142 MeV, $\alpha$=0.15 fm$^{-2}$.

They also include two forbidden bound states which corresponded to the schemes {52} and {7}. The phase shifts for these potentials are shown in Fig. 13 by the dashed and dot-dashed lines in comparison with the results of our phase shift analysis given by black points and triangles.

The potential of the doublet $^2P$-wave of scattering can be represented, for example, by the

next parameters:

$^2P$ - $V_0$=-68.0 MeV, $\alpha$=0.1 fm$^{-2}$.

The results of the calculation of the phase shifts with this potential are shown in Fig. 14 by the solid line. The potential has one forbidden bound state with the scheme {61} and the allowed state with the Young's schemes {43} and {421}.

Such potential gives the wrong bound energy of the $^7$Be nucleus in the p$^6$Li channel because the allowed state is mixed in terms of the above mentioned symmetries, but only the scheme {43} corresponds to the ground bound state [64]. But even if we use the methods of receiving the "pure" phase shifts given in [64], it is not possible to obtain the "pure" in Young's schemes potential of the ground state. It seams that it is due to the absence of the spin-orbital splitting and the small probability of clusterization of the $^7$Be nucleus into the p$^6$Li channel.

That is why the "pure" according to orbital symmetries $^2P_{3/2}$-potential of the ground state of the $^7$Be nucleus with Young's scheme {43} should be constructed so as to describe primarily the channel energy - the bound energy of the ground state of the nucleus with J=3/2$^-$ as the p$^6$Li system and its root-mean-square radius. Then the parameters of the "pure" $^2P_{3/2}^{\{43\}}$-potential can be represented as

$^2P_{3/2}$ - $V_P$=-252.914744 MeV, $\alpha_P$=0.25 fm$^{-2}$. (21)

Such potential gives the bound energy of the allowed state with scheme {43} equal to -5.605800 MeV, while the experimental value is equal to -5.6058 MeV [65] and has one forbidden state corresponding to Young's scheme {61}. The root-mean-square charge radius is equal to 2.63 fm what is generally in agreement with the data of [65], and the $C_W$ constant (8) is equal to 2.66(1) within the range of 5-13 fm.

For the parameters of the $^2P_{1/2}^{\{43\}}$-potential of the first forbidden state of $^7$Be nucleus with J=1/2$^-$ the next values are obtained

$^2P_{1/2}$ - $V_P$=-251.029127 MeV, $\alpha_P$=0.25 fm$^{-2}$. (22)

This potential leads to the bound energy -5.176700 MeV while its experimental value is equal to -5.1767 MeV [65] and it contains the forbidden state with scheme {61}. The asymptotic constant (8) is equal to 2.53(1) within the range of 5-13 fm and the charge radius is equal to 2.64 fm. The absolute accuracy of searching for the bound energy in our new computer programs was taken to be at the level of $10^{-7}$ MeV.

The obtained potential parameters of the bound states a little bit differ from our previous results [64]. This is because we used in these calculations the exact mass values of particles and more accurate description of the experimental values of the energy levels.

The variational method for the energy of the ground state gives the value -5.605797 MeV and hence the average energy for this potential is equal to -5.6057985(15) MeV, i.e. the accuracy of its determination equals ±1.5 eV. The asymptotic constant at the distances of 5-13 fm turned out to be comparatively stable and equal to 2.67(2) and the charge radius is in agreement with the calculation results based on the finite-difference method. The variational wave function (7) for the ground state of the $^7$Be nucleus in the p$^6$Li channel with potential (21) is listed in Table 9, and the residual error is not more than $10^{-12}$.

Table 9
The variational parameters and expansion coefficients of the radial WF of the bound state of the p$^6$Li system for the $P$-potential (21). The normalization of the function with these coefficients in the range 0-25 fm equals N=0.9999999999999895

| $i$ | $\beta_i$ | $C_i$ |
| --- | --- | --- |
| 1 | 2.477181344627947E-002 | 1.315463702527344E-003 |
| 2 | 5.874061769072439E-002 | 1.819913407984276E-002 |
| 3 | 1.277190608958812E-001 | 9.837541674753882E-002 |
| 4 | 2.556552559403827E-001 | 3.090018297080802E-001 |
| 5 | 6.962545656024610E-001 | -1.195304944694753 |
| 6 | 87.215179556255360 | 3.237908749007494E-003 |
| 7 | 20.660304078047520 | 5.006096657700867E-003 |
| 8 | 1.037788131786810 | -6.280751485496025E-001 |
| 9 | 2.768782138965186 | 1.282309968994793E-002 |
| 10 | 6.753591325944827 | 8.152343478073063E-003 |

For the energy of the first excited state with the use of the VM we received the value of -5.176697 MeV and hence the average energy is equal to -5.1766985(15) MeV, with the same accuracy as it was in the case of the GS. The asymptotic constant at distances of 5-13 fm turned out to be of the level of 2.53(2), the residual error being not more than 10$^{-12}$ and the charge radius being almost the same as for the GS. The parameters of the excited state of the WF of the $^7$Be nucleus with potential (22) are listed in Table 10.

Table 10
The variational parameters and expansion coefficients of the radial WF of the first excited bound state of the p$^6$Li system for the $P$-potential (22). The normalization of the function with these coefficients in the range 0-25 fm equals N=0.9999999999999462

| $i$ | $\beta_i$ | $C_i$ |
| --- | --- | --- |
| 1 | 2.337027900191992E-002 | 1.218101547601343E-003 |
| 2 | 5.560733180673633E-002 | 1.653319276756672E-002 |
| 3 | 1.214721917930904E-001 | 9.009619752334307E-002 |
| 4 | 2.474544878067495E-001 | 3.003291466882630E-001 |
| 5 | 7.132725465249825E-001 | -1.332325501226168 |
| 6 | 84.896023494945160 | 3.273725679869025E-003 |
| 7 | 1.162854732120233 | -5.340018423135894E-001 |
| 8 | 1.574203000936825 | 9.367648737801053E-002 |
| 9 | 5.779896847077723 | 1.033713941440747E-002 |
| 10 | 19.422905786572090 | 5.314592946045428E-003 |

It should be noted that on the basis of the obtained results of the phase shift analysis for the doublet $^2P$-phase shift of scattering shown in Fig. 14, it is impossible to construct a unique $^2P$-potential. The results of the phase shift analysis at higher energies are required and they have to take into account $^2P$-wave and spin-orbital phase shift splitting.

The same concerns the $^4S$-potential, and only the $^2S$-interaction is obtained quite uniquely. This interaction with the above $^2P$-potentials of the bound states can be used in future, for example, for the calculations of the astrophysical $S$-factor with the $E$1 transition from the doublet $^2S$-wave of scattering to the ground and first excited doublet bound $^2P$-states of the $^7$Be nucleus.

*5.3. Astrophysical S-factor*

Although the p$^6$Li reaction of radiative capture may be of some interest for the nuclear astrophysics [66], it is not well enough studied experimentally. There is a comparatively small number of measurements of the total cross-sections and calculations of the *S*-factor [9], and they were performed only in the energy range from 35 keV to 1.2 MeV. Nevertheless, it would be interesting to consider the possibility of its description in the frame of the potential cluster model taking into account the classification of the bound states according to the orbital Young's schemes [64,67] at the astrophysical energy range where the experimental data exist.

The *E*1 transitions from $^2S$ and $^2D$-states of scattering to the ground $^2P_{3/2}$ and the first excited $^2P_{1/2}$ bound states of the $^7$Be nucleus were taken into account when the astrophysical *S*-factor was considered. The calculation of the wave function of the $^2D$-wave without spin-orbital splitting was made on the basis of the $^2S$-potential but with the orbital momentum *L*=2.

When the calculations were made it turned out that the given above $^2S$-potential of scattering with the depth of 110 MeV and based on the phase shift analysis [63] led to the astrophysical *S*-factor significantly lower than it had to be. At the same time the doublet $^2S$-potential with the depth of 126 MeV, which was obtained in our calculations, gives quite correct description of the general behavior of the experimental *S*-factor. The results received are shown in Fig. 15. The results of the transitions from $^2S$ and $^2D$-waves of scattering to the ground state of the $^7$Be nucleus are shown by the dashed line, the dotted line is for the transitions to the first excited state and the solid line is the total *S*-factor. Black points, triangles and circles are the experimental data from works [68] given in work [69].

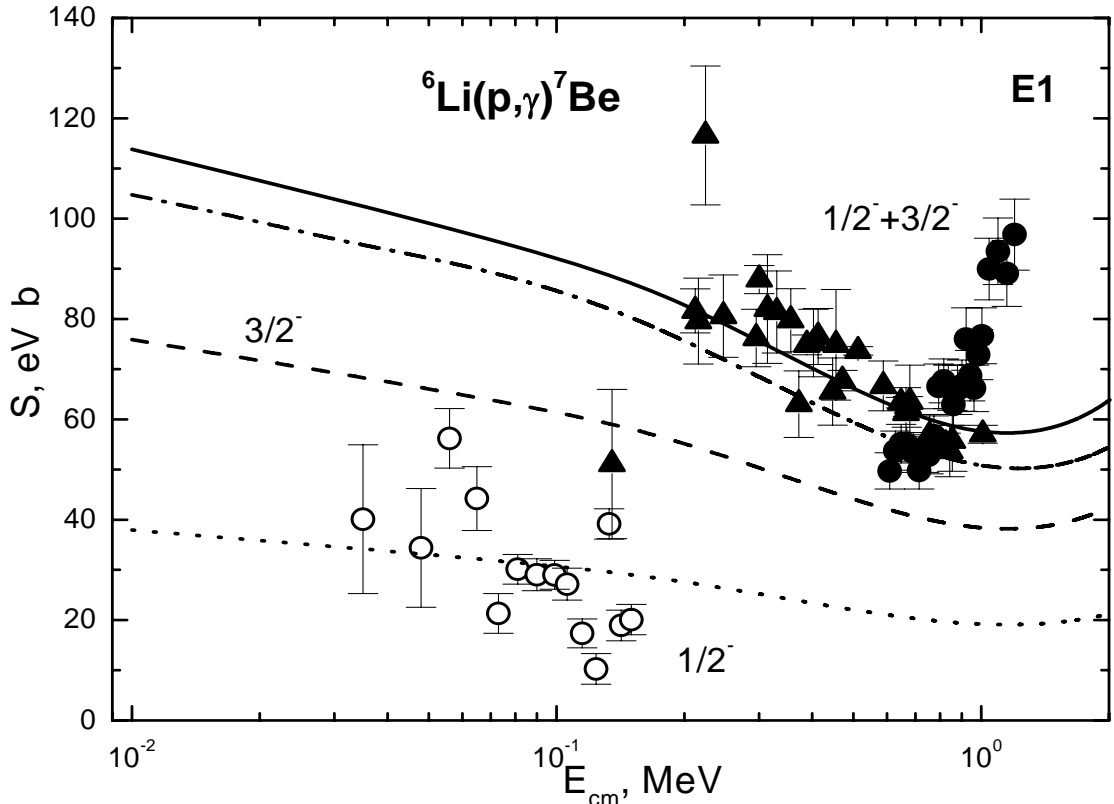

Fig. 15. Astrophysical *S*-factor of p$^6$Li radiative capture. Black points, triangles and circles are the experimental data from works [68] given in work [69]. The result for transitions from $^2S$ and $^2D$-waves of scattering to the ground state of the $^7$Be nucleus is shown by the dashed line, and for transitions to the first excited state - by the dotted line. The solid line shows the total *S*-factor.

The calculated *S*-factor at 10 keV is equal to $S(3/2^-)$=76 eV b and $S(1/2^-)$=38 eV b while the total value is equal to 114 eV b. The behavior of the $S(1/2^-)$-factor well describes the experimental data (circles in Fig.15) for the transition to the first excited state of the $^7$Be nucleus at low energies. Since the calculated $S$-factor at 10 keV continues to grow up, it is difficult to make any conclusions about its value at zero energy.

For comparison of the calculated $S$-factor at zero energy (10 keV) we will give the known results for the total $S(0)$: 79(18) eV b [70], 105 eV b (at 10 keV) [69] and 106 eV b [71]. In work [72] for the $S$-factor of transitions to the ground state the value of 39 eV b is given and for the transition to the first excited state the value of 26 eV b, so the total $S$-factor is equal to 65 eV b. As it seen the difference between these data is comparatively large, and our results are in agreement with them in general.

Besides, a small change in the depth of the $^2S$-potential of scattering, for example if we take 124 MeV, which practically does not affect the behavior of the calculated phase shifts shown in Fig. 13 by the short dashes, influences the $S$-factor significantly and leads to the value of 105 eV b at 10 keV. The total $S$-factor with this potential is shown in Fig. 15 by the dot-dashed line which lies within the experimental error band at the energies below 1 MeV.

It should be mentioned that if we use the potentials without the forbidden states in $S$ and $P$-waves or with another number of FS, then the value of the calculated $S$-factor turns out to be from 3 to 100 times lower that the values obtained above. For example, the $^2S$-potential with one forbidden state and parameters 25 MeV and 0.15 fm$^{-2}$, which gives a good description of the phase shifts of scattering and the given above potential of the ground state, leads to the $S$-factor of about 1 eV b.

Thus, the doublet $^2S$-phase shifts obtained in our phase shift analysis, which takes into account the doublet $^2P$-phase shift, lead to the potential which allowed to describe the experimental $S$-factor at the energies down to 1 MeV, in distinction from the interaction constructed on the basis of the analysis results [63]. The potential cluster model used and the potentials given above allow in general obtaining quite reasonable results for the description of the process of radiative p$^6$Li capture at the astrophysical energy range, as it was in the case of lighter nuclei [73].

# 6. p$^{12}$C radiative capture

In this section we will consider the p$^{12}$C system and the process of proton radiative capture by the $^{12}$C nucleus at astrophysical energies. The new measurement of differential cross-sections of the elastic p$^{12}$C scattering at energies from 200 keV up to 1.1 MeV (center-of-mass system) within the range of 10$^0$-170$^0$ with 10% errors was carried out in works [4].

Further, the standard phase shift analysis made and the potential of S-state of p$^{12}$C system was reconstructed in this paper on the basis of these measurements [39], and then the astrophysical S-factor at the energies down to 20 keV was considered in the frame of potential cluster model.

*6.1. Phase shift analysis*

While examining scattering in the particle system with a total spin 1/2, i.e. where one of the particles has spin 0, and the second has spin 1/2, it is necessary to take into account spin-orbital splitting of phase shifts. This sort of scattering takes place in the nuclear systems such as N$^4$He, $^3$H$^4$He, p$^{12}$C etc. The differential cross-section of the elastic scattering of nuclear particles is represented as [62]

$$\frac{d\sigma(\theta)}{d\Omega} = |A(\theta)|^2 + |B(\theta)|^2.$$

The connection between the differential cross-sections of the elastic scattering and the phase shifts is given in work [62] too.

Earlier, the phase shift analysis of excitation functions of the p$^{12}$C scattering, measured in [74] at energies 400-1300 keV (l.s.-laboratory system) and angles from 106$^0$ to 169$^0$, was carried out in work [75] where it was found that the S-phase must be in the range of 153$^0$-154$^0$ at the energy $E_{lab}$=900 keV.

For the same experimental data we have received 152.7$^0$. The cross-sections were extracted from the excitation functions [75] at energies 866-900 keV. Results of our calculations $\sigma_t$ in comparison with experimental data $\sigma_e$ are given in Table 11. Partial $\chi^2_i$ for each point with 10% errors in experimental cross-sections are given in the last column of the table and the value of 0.11 is received for the average of $\chi^2$.

Table 11
Comparison of theoretical and experimental cross-sections of the elastic p$^{12}$C scattering at the energy 900 keV

| $\theta^0$ | $\sigma_e$, (mb) | $\sigma_t$, (mb) | $\chi^2_i$ |
|---|---|---|---|
| 106 | 341 | 341.5 | 1.90E-04 |
| 127 | 280 | 282.1 | 5.76E-03 |
| 148 | 241 | 251.2 | 1.80E-01 |
| 169 | 250 | 237.5 | 2.50E-01 |

The values 155$^0$-157$^0$ of the S-phase shift are found in work [75] at the energy 751 keV (l.s.). The received results for this energy are listed in Table 12. We took the cross-section data from excitation functions at energies in the range of 749-754 keV and obtained the value 156.8$^0$ for the S-phase shift at the average value of $\chi^2$=0.30.

Table 12
Comparison of theoretical and experimental cross-sections of the elastic p$^{12}$C scattering at the energy 750 keV

| $\theta^0$ | $\sigma_e$, (mb) | $\sigma_t$, (mb) | $\chi^2_i$ |
|---|---|---|---|
| 106 | 428 | 428.3 | 3.44E-05 |
| 127 | 334 | 342.8 | 6.91E-02 |
| 148 | 282 | 299.1 | 3.66E-01 |
| 169 | 307 | 279.9 | 7.82E-01 |

All these reference results are in a good agreement with each other and the phase shift analysis [39] of the new experimental data of the differential cross-section of p$^{12}$C scattering was carried out using our program at energies in the range of 230-1200 keV (l.s.) [4]. The results of our phase shift analysis are given in Table 13 and are shown in Fig. 16 in comparison with the values of work [75] which are shown by the dashed line.

Fig. 17a, b, c shows the differential cross-sections in the resonance region at 457 keV (l.s.), the calculation results of these cross-sections on the basis of Rutherford formula - (dotted line), the cross-sections on the basis of our phase shift analysis which takes into account *S*-phase shift only (solid line). The cross-sections which take into account the *S* and *P*-waves of scattering in the phase shift analysis are shown by the dashed line.

Table 13
Results of the phase shift analysis of the elastic p$^{12}$C scattering at low energies taking into account S-wave only

| $E_{cm}$, (keV) | $S_{1/2}$, (deg.) | $\chi^2$ |
|---|---|---|
| 213 | 2.0 | 1.35 |
| 317 | 2.5 | 0.31 |
| 371 | 7.2 | 0.51 |
| 409 | 36.2 | 0.98 |
| 422 | 58.0 | 3.75 |
| 434 | 107.8 | 0.78 |
| 478 | 153.3 | 2.56 |
| 689 | 156.3 | 2.79 |
| 900 | 153.6 | 2.55 |
| 1110 | 149.9 | 1.77 |

One can see from the figures that it is impossible to describe the cross-section in the resonance region on the basis of *S*-wave only. In this case, the *P*-wave shown in Fig. 18 starts playing a considerable role and its consideration improves the experiment description. At the energy 457 keV (l.s.) the value of $\chi^2$.can be improved from 3.75 to 0.79 (Fig. 17b) by taking into account *P*-wave, what considerably affects the quality of description of differential cross-sections.

It is seen in Fig. 18, that at low energies $P_{1/2}$-phase shift lies above $P_{3/2}$ but at the energy about 1.2 MeV they intersect and then $P_{3/2}$-phase shift lies above in the region of negative angles [76,77]. The value of the *S*-phase shift practically doesn't change when the *P*-wave is taken into account (see Fig. 16 - blank blocks). Consideration of the *D*-wave in the phase shift analysis leads to the value 1-1.5 degrees in the resonance region and practically doesn't affect the behavior of calculated differential cross-sections.

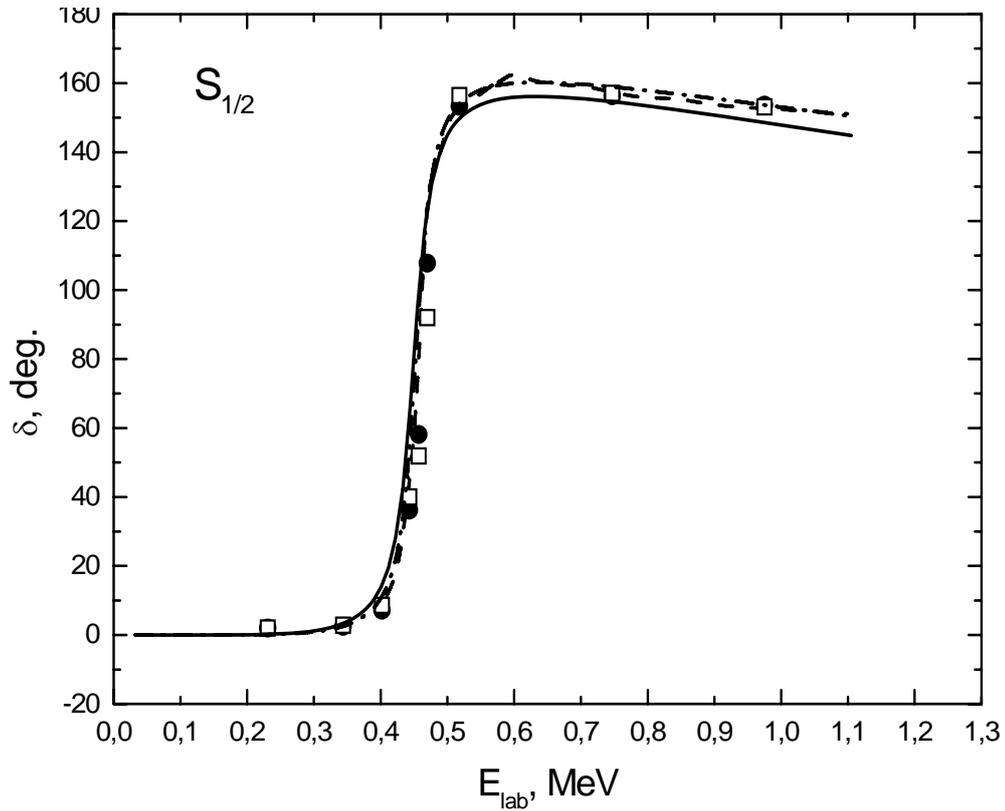

Fig. 16. $^2S$-phase shift of the elastic p$^{12}$C scattering at low energies. Black points - results of the phase shift analysis for the S-phase shift taking into account the S-wave only; blank blocks - results of the phase shift analysis for the S-phase shift taking into account S and P-waves; dotted line - results of work [75]; other lines - results calculated with different potentials.

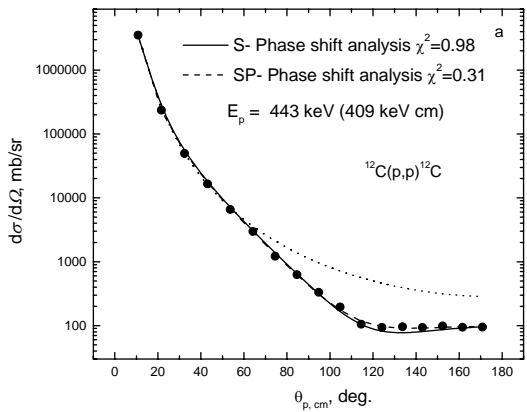

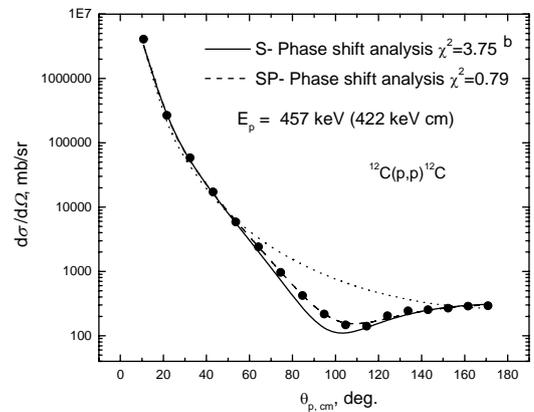

Figs. 17a, b, c. Differential cross-sections of p$^{12}$C scattering. Solid line is the phase shift analysis which takes into account the *S*-wave only; dotted line - the Rutherford scattering; dashed line - the phase shift analysis where *S* and *P*-waves are taken into account; black points - experimental data [4].

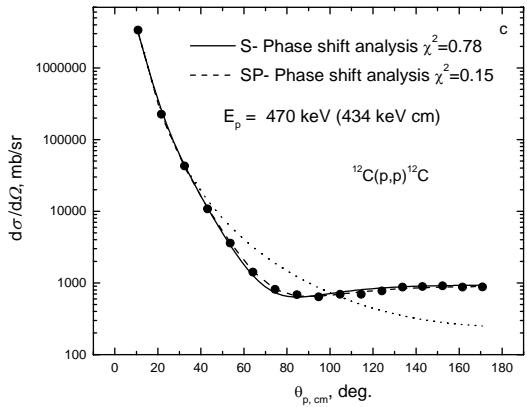

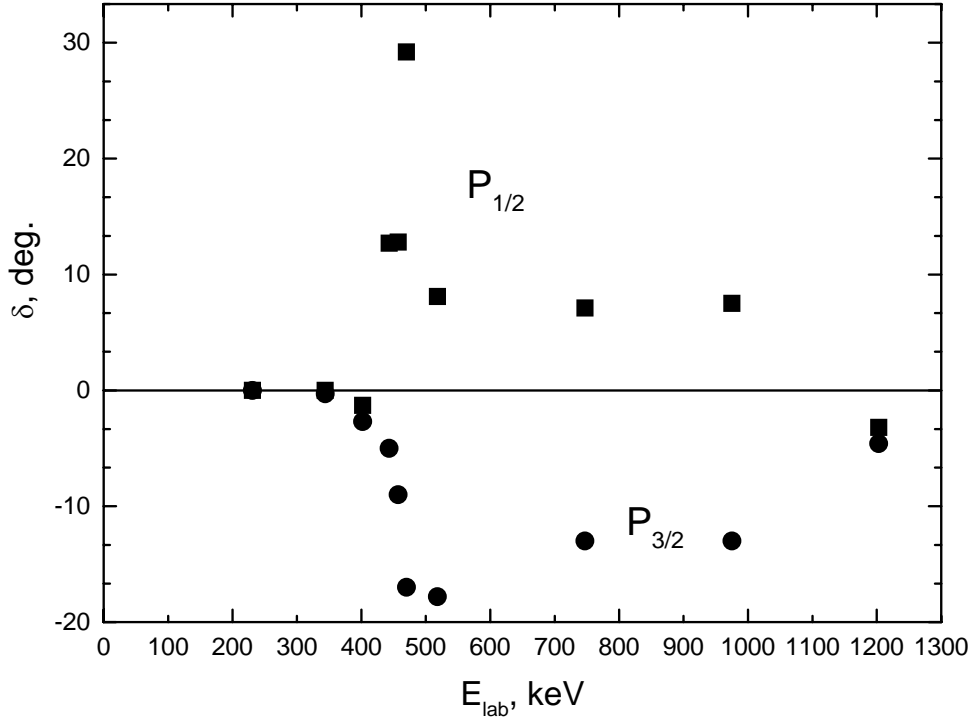

Fig. 18. $^2P$-phase shifts of the elastic p$^{12}$C scattering. Black points ($P_{3/2}$) and blocks ($P_{1/2}$) are the results of the phase shift analysis for $P$-phase shifts taking into account $S$ and $P$-waves.

*6.2. Astrophysical S-factor*

The radiative p$^{12}$C capture at low energies is the part of the CNO thermonuclear cycle and gives a considerable contribution into the energy output of thermonuclear reactions [15,66]. The existing experimental data of the astrophysical *S*-factor [9] indicates the presence of the narrow resonance with the width of about 32 keV at the energy 0.422 MeV (center-of-mass system), which leads to the two-three order increase in the *S*-factor.

It is interesting to find out if there is a possibility to describe the resonance *S*-factor on the basis of the PCM with FS and with the classification of orbital states according to Young's schemes. The phase shift analysis of the new experimental data [4] of differential cross-sections of the elastic p$^{12}$C scattering at astrophysical energies [39], which we have shown above, allows constructing the potentials of the p$^{12}$C interaction for the phase shift analysis of the elastic scattering.

The $E1(L)$ transition resulting from the orbital part of electric $Q_{JM}(L)$ operator [10] is taken into account in present calculations of the process of radiative p$^{12}$C capture. The cross-sections of $E2(L)$ and $MJ(L)$ processes and the cross-sections depending on the spin part $EJ(S)$, $M2(S)$ turned out to be a few orders less. The electrical $E1(L)$ transition in the p$^{12}$C→γ$^{13}$N process is possible between the doublet $^2S_{1/2}$ and $^2D_{3/2}$-states of scattering and the ground bound $^2P_{1/2}$-state of the $^{13}$N nucleus in the p$^{12}$C channel.

Let's examine the classification of orbital states according to Young's schemes in the p$^{12}$C system for the purposes of construction of the interaction potential. The possible orbital Young's schemes in the $N=n_1+n_2$ particle system can be defined as the direct external product of orbital schemes of each subsystem, in our case it gives {1}×{444}={544} and {4441} [78]. The first of them is consistent only with the orbital momentum $L=0$ and is forbidden, because s-shell cannot contain more than four nucleons. The second scheme is allowed with orbital momenta 1 and 3 [78], the first of which corresponds to the ground bound state of the $^{13}$N nucleus with $J=1/2^-$. Therefore, the forbidden BS has to be in the potential of $^2S$-wave, and $^2P$-wave has an allowed state only at the energy of -1.9435 MeV [79].

For the calculations of photonuclear processes the nuclear part of the inter-cluster p$^{12}$C interaction is represented as (6) with the pointlike Coulomb component. The potential of $^2S_{1/2}$-wave is constructed so as to describe correctly the corresponding partial phase shift of the elastic scattering, which has a well defined resonance at 0.457 MeV (l.s.).

Using the results of the phase shift analysis [39] the $^2S_{1/2}$-potential of the p$^{12}$C interaction with FS at energy $E_{FS}$=-25.5 MeV was obtained together with parameters:

$V_S$ = -67.75 MeV, $\alpha_S$ = 0.125 fm$^{-2}$.

The results of calculation of $^2S_{1/2}$-phase shift with this potential are shown in Fig. 16 by the solid line.

The potential of the bound $^2P_{1/2}$-state has to reproduce correctly the bound energy of the $^{13}$N nucleus in the p$^{12}$C channel -1.9435 MeV [79] and reasonably describe the root-mean-square radius, which probably does not differ significantly from the radius of the $^{14}$N nucleus equal to 2.560(11) fm [79]. As a result the following parameters were received:

$V_{GS}$ = -81.698725 MeV, $\alpha_{GS}$ = 0.22 fm$^{-2}$. (23)

The potential gives the bound energy equal to -1.943500 MeV and the root-mean-square radius $R_{ch}$=2.54 fm. We use the following values for the radii of proton and $^{12}$C: 0.8768(69) fm [12] and 2.472(15) fm [80]. The asymptotic constant $C_W$ with Whittaker asymptotics (8) was calculated for controlling behavior of WF of BS at long distances; its value in the range of 5-20 fm equals 1.96(1).

The results of calculations of the S-factor of the radiative p$^{12}$C capture with the abovementioned potentials of $^2P_{1/2}$ and $^2S_{1/2}$-waves at energies from 20 keV to 1.0 MeV are shown in Fig. 19 by the solid line and the experimental data are taken from review [9]. The value 3.0 keV b of the S-factor is received at energy 25 keV. The extrapolation of the S-factor experimental measurements gives: 1.45(20) keV b and $1.54^{+15}_{-10}$ keV b [79]. The $^2S$-potential given here is not the only one which can describe the resonance behavior of the S-phase shift at energies lower than 1 MeV.

Thus, it is always possible to find other combinations of bound and scattering state potentials with FS which lead to the similar results for the $^2S_{1/2}$-phase shift and describe well the value and location of the maximum of the S-factor, for example:

$V_{GS}$=-121.788933 MeV, $\alpha_{GS}$=0.35 fm$^{-2}$,
$R_{ch}$=2.49 fm, $C_W$=1.50(1), $E_{GS}$=-1.943500 MeV,
$V_S$=-102.05 MeV, $\alpha_S$=0.195 fm$^{-2}$, $E_{FS}$=-12.8 MeV. (24)

They lead to a sharper fall of the S-factor at energies near the resonance. The phase shift of potential (24) and the behavior of its S-factor are shown in Figs. 16 and 19 by dot-dashed lines. The S-factor value for this combination of the potentials at 25 keV is equal to 1.85 keV b, what generally agrees with the values given in review [79].

A narrower bound state potential with the same potential of scattering (24)

$V_{GS}$=-144.492278 MeV, $\alpha_{GS}$=0.425 fm$^{-2}$,
$R_{ch}$=2.47 fm, $C_W$=1.36(1), $E_{GS}$=-1.943500 MeV, (25)

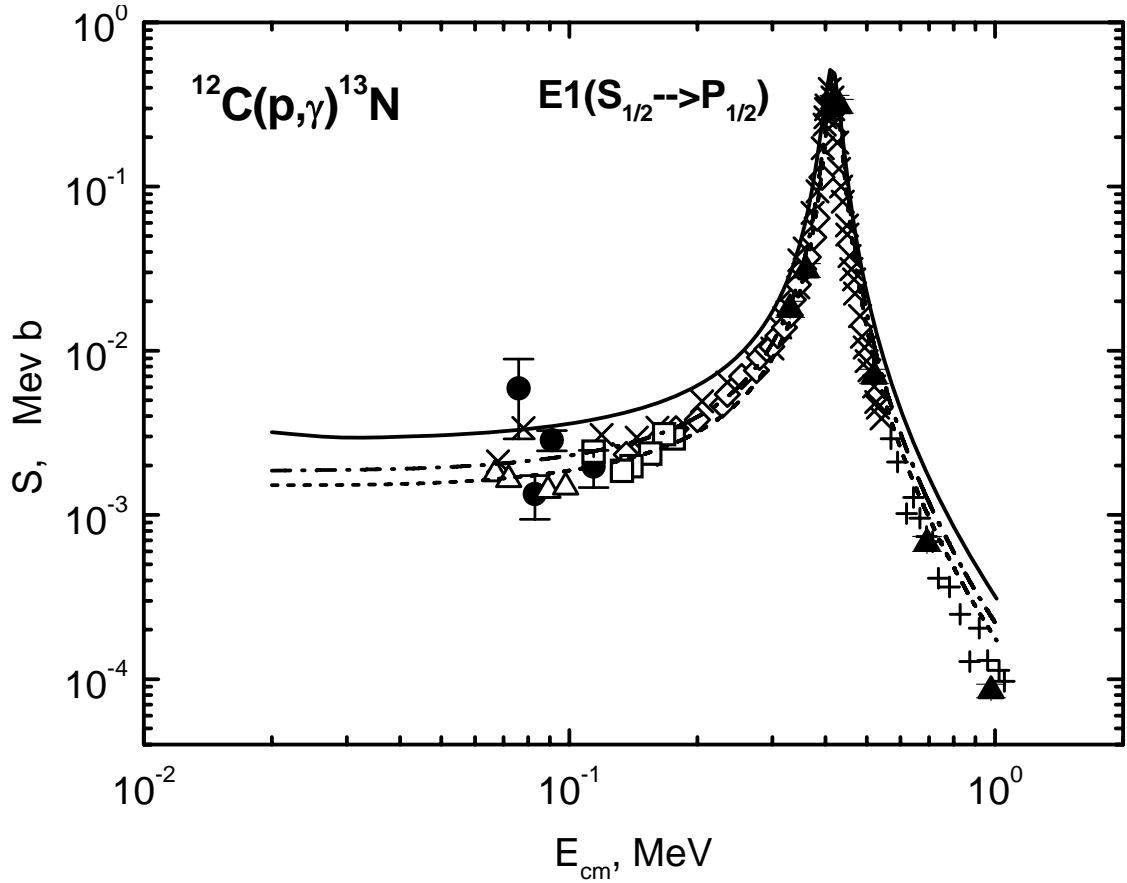

Fig. 19. Astrophysical *S*-factor of p$^{12}$C radiative capture at low energies. The experimental data specified as ×, •, □, + and Δ are taken from review [9], triangles are from [81]. Lines: calculations with different potentials.

leads to a small decrease in the *S*-factor at the resonance energy, as it is shown in Fig. 19 by the dashed line, and gives the value *S*(25)=1.52 keV b which is in a good agreement with data [79]. At the same time, in the range of 20-30 keV the *S*-factor value is practically constant and one can consider it as the *S*-factor value at zero energy with an error of about 0.02 keV b. As it can be seen from the above results, the asymptotic constant and charge radius of the nucleus became smaller as the width of the BS potentials decreases, and potential (25) gives their smallest values.

The variational method was used for an additional control of the accuracy of bound energy calculations, which allowed to obtain the bound energy of -1.943498 MeV for the first variant of the potential (23) by using an independent variation of parameters and the grid having dimension 10. The asymptotic constant $C_W$ of the variational WF at distances of 5-20 fm remains at the level of 1.97(2). Its variational parameters are listed in Table 14. The charge radius does not differ from the value obtained in FDM calculations.

For the real bound energy in this potential it is possible to use the value -1.943499(1) MeV, i.e. the calculation error of finding bound energy is on the level of ±1 eV, because the variational energy decreases as the dimension of the basis increases and gives the upper limit of the true bound energy, but the finite-difference energy increases as the size of steps decreases and the number of steps increases.

The variant (24) of the BS potential, which was examined within the frame of the variational method, leads to the bound energy of -1.943498 MeV with the residual error of the order of $3 \cdot 10^{-14}$, the radius equal to 2.49 fm and the asymptotic constant at distances of 5-17 fm equal to 1.50(2). The variational parameters and expansion coefficients of the radial wave function are listed in Table 15.

Table 14
The variational parameters and expansion coefficients of the radial WF of the p$^{12}$C system for the first variant (23) of the BS potential

| $i$ | $\beta_i$ | $C_i$ |
|---|---|---|
| 1 | 4.310731038130567E-001 | -2.059674967002619E-001 |
| 2 | 1.110252143696502E-002 | -1.539976053334172E-004 |
| 3 | 4.617318488940146E-003 | -2.292772895754105E-006 |
| 4 | 5.244199809745243E-002 | -1.240687319547592E-002 |
| 5 | 2.431248255158095E-002 | -1.909626327101099E-003 |
| 6 | 8.481652230536312 E-000 | 5.823965673819461E-003 |
| 7 | 1.121588023402944E-001 | -5.725546189065398E-002 |
| 8 | 2.309223399000618E-001 | -1.886468874357471E-001 |
| 9 | 2.297327380843046 E-000 | 1.244238759439573E-002 |
| 10 | 3.7567721497435540 E+001 | 3.435757447077250E-003 |

Table 15
The variational parameters and expansion coefficients of the radial WF of the p$^{12}$C system for the second variant (24) of the BS potential

| $i$ | $\beta_i$ | $C_i$ |
|---|---|---|
| 1 | 1.393662782203888E-002 | 3.536427343510346E-004 |
| 2 | 1.041704259743847E-001 | 3.075071412877344E-002 |
| 3 | 4.068236340341411E-001 | 3.364496084003433E-001 |
| 4 | 3.517787678267637E-002 | 4.039427231852849E-003 |
| 5 | 2.074448420678197E-001 | 1.284484754736406E-001 |
| 6 | 7.360025091178769E-001 | 2.785322894825304E-001 |
| 7 | 3.551046173695889E-000 | -1.636661944722212E-002 |
| 8 | 1.5131407009411240E+001 | -9.289494991217288E-003 |
| 9 | 9.726024028584802E-001 | -1.594107798542716E-002 |
| 10 | 6.634603967502104E-002 | 8.648073851532037E-003 |

The third variant (25) of the BS potential, which was examined within the frame of variational method, leads to the bound energy of -1.943499 MeV with the residual error of 6·10$^{-14}$, the radius being the same as in the FDM calculations and the asymptotic constant at distances of 5-17 fm being equal to 1.36(2). The variational parameters of the radial wave function are listed in Table 16.

Thus, the given above pairs of the potentials with FS for the $^2S_{1/2}$-wave and the bound state, which gives the correct bound energy, lead to the joint description of the resonance in the $S$-factor and the resonance in the $^2S$-phase of scattering. The best results of the description of the scattering process characteristics of the radiative p$^{12}$C capture at low energies and of the $S$-factor at 25 keV are obtained with the third variant (25) of the interaction potential.

At the same time, if we use the potentials of the $^2S_{1/2}$-wave with small depth and without forbidden states, for example, with parameters:

$V_S$=-15.87 MeV, $\alpha_S$=0.1 fm$^{-2}$,
$V_S$=-18.95 MeV, $\alpha_S$=0.125 fm$^{-2}$, (26)
$V_S$=-21.91 MeV, $\alpha_S$=0.15 fm$^{-2}$,

Table 16
The variational parameters and expansion coefficients of the radial WF of the p$^{12}$C system for the third variant (25) of the BS potential

| $i$ | $\beta_i$ | $C_i$ |
|---|---|---|
| 1 | 1.271482702554672E-002 | 2.219877609724907E-004 |
| 2 | 9.284155511162226E-002 | 2.240043561912315E-002 |
| 3 | 3.485413978134982E-001 | 2.407314126671507E-001 |
| 4 | 3.088717918378341E-002 | 2.494885124596691E-003 |
| 5 | 1.815363020074388E-001 | 8.792233462610707E-002 |
| 6 | 5.918532693855678E-001 | 3.652121068403727E-001 |
| 7 | 3.909887088341156E+000 | -1.906081640167417E-002 |
| 8 | 1.6356080812096500 E+001 | -1.111922033874987E-002 |
| 9 | 9.358886757095011E-001 | 2.314583156796476E-001 |
| 10 | 5.673177540516311E-002 | 5.956470542991426E-003 |

then we can't obtain the correct description of the maximum of the *S*-factor of the radiative capture. It is impossible to describe the absolute value of the *S*-factor which for all variants of the scattering potentials (26) and the BS potentials is 2-3 times as much as the experimental maximum. At the same time, for all given depthless potentials of the form (26) the resonance behavior of the $^2S_{1/2}$-phase shift of scattering is well described. As the width of the $^2S_{1/2}$-potential decreases, i.e. the α value increases, the value of the *S*-factor maximum grows up, e.g. for the last variant of the $^2S_{1/2}$-scattering potential its value is approximately three times as much as the experimental value.

It should be noted that in all calculations the cross-section of the *E*1 electrical process due to transition from the doublet $^2D_{3/2}$-state of scattering to the ground bound $^2P_{1/2}$-state of the $^{13}$N nucleus is 4-5 orders less than the cross-section of the transition from $^2S_{1/2}$-state of scattering. Thus, the main contribution to the calculated *S*-factor of the p$^{12}$C→$^{13}$Nγ process is made by the *E*1 transition from the $^2S$-wave of scattering to the ground state of the $^{13}$N nucleus. The mass of proton was taken to be 1 in all calculations for the p$^{12}$C system.

Thus, it is possible to combine the description of the astrophysical *S*-factor and the $^2S_{1/2}$-phase shift in the resonance energy range 0.457 MeV (l.s.) on the basis of the PCM and the deep $^2S_{1/2}$-potential with the FS, and to receive the reasonable values for the charge radius and asymptotic constant. The depthless potentials of scattering do not lead to the joint description of the *S*-factor and the $^2S$-phase shift of scattering at any considered combinations of p$^{12}$C interactions [82].

## 7. Conclusion

The description of behavior of the *S*-factors in all considered systems at low energies may be viewed as a certain evidence in favor of the potential approach in cluster model. The inter-cluster interactions including FS are constructed on the basis of the phase shifts of the cluster elastic scattering, and each partial wave is described by its potential, for example of the Gaussian form, with certain parameters.

The splitting of the general interaction into the partial waves allows detailing its structure and the classification of the orbital states according to Young's schemes allows identifying the presence and the number of the forbidden states. It gives the possibility to find the number of nodes of the WF of cluster relative motion and leads to a definite depth of the interaction allowing to avoid the discrete ambiguity of the potential depth as it is the case in the optical model.

The form of each partial phase shift of scattering can be correctly described only at a certain width of such a potential which saves us from the continuous ambiguity also characteristic of the well-known optical model. As a result, all the parameters of such a potential are fixed quite uniquely, and the "pure" according to Young's schemes interaction component allows describing the basic characteristics of the bound state of the lightest clusters correctly, which is realized in the light atomic nuclei with a high probability.

However, all the above-mentioned is correct provided that the phase shifts of scattering are obtained correctly from the experimental data of the elastic scattering. Up to present, for the majority of the lightest nuclear systems the phase shifts of scattering have been received with rather big errors, sometimes reaching 20-30%. This makes the construction of the exact potentials of the inter-cluster interaction very difficult and, finally, leads to significant ambiguities in the final results obtained in the potential cluster model.